\documentclass{mn2e}
\usepackage{epsfig}
\usepackage{psfig}
\usepackage{rotating}
\newread\epsffilein    
\newif\ifepsffileok    
\newif\ifepsfbbfound   
\newif\ifepsfverbose   
\newdimen\epsfxsize    
\newdimen\epsfysize    
\newdimen\epsftsize    
\newdimen\epsfrsize    
\newdimen\epsftmp      
\newdimen\pspoints     
\def\epsfbox#1{\global\def\epsfllx{72}\global\def\epsflly{72}%
   \global\def\epsfurx{540}\global\def\epsfury{720}%
   \def\lbracket{[}\def\testit{#1}\ifx\testit\lbracket
   \let\next=\epsfgetlitbb\else\let\next=\epsfnormal\fi\next{#1}}%
\def\epsfgetlitbb#1#2 #3 #4 #5]#6{\epsfgrab #2 #3 #4 #5 .\\%
   \epsfsetgraph{#6}}%
\def\epsfnormal#1{\epsfgetbb{#1}\epsfsetgraph{#1}}%
\def\epsfgetbb#1{%
\openin\epsffilein=#1
\ifeof\epsffilein\errmessage{I couldn't open #1, will ignore it}\else
   {\epsffileoktrue \chardef\other=12
    \def\do##1{\catcode`##1=\other}\dospecials \catcode`\ =10
    \loop
       \read\epsffilein to \epsffileline
       \ifeof\epsffilein\epsffileokfalse\else
          \expandafter\epsfaux\epsffileline:. \\%
       \fi
   \ifepsffileok\repeat
   \ifepsfbbfound\else
    \ifepsfverbose\message{No bounding box comment in #1; using defaults}\fi\fi
   }\closein\epsffilein\fi}%
\def\epsfsetgraph#1{%
   \epsfrsize=\epsfury\pspoints
   \advance\epsfrsize by-\epsflly\pspoints
   \epsftsize=\epsfurx\pspoints
   \advance\epsftsize by-\epsfllx\pspoints
   \epsfxsize\epsfsize\epsftsize\epsfrsize
   \ifnum\epsfxsize=0 \ifnum\epsfysize=0
      \epsfxsize=\epsftsize \epsfysize=\epsfrsize
     \else\epsftmp=\epsftsize \divide\epsftmp\epsfrsize
       \epsfxsize=\epsfysize \multiply\epsfxsize\epsftmp
       \multiply\epsftmp\epsfrsize \advance\epsftsize-\epsftmp
       \epsftmp=\epsfysize
       \loop \advance\epsftsize\epsftsize \divide\epsftmp 2
       \ifnum\epsftmp>0
          \ifnum\epsftsize<\epsfrsize\else
             \advance\epsftsize-\epsfrsize \advance\epsfxsize\epsftmp \fi
       \repeat
     \fi
   \else\epsftmp=\epsfrsize \divide\epsftmp\epsftsize
     \epsfysize=\epsfxsize \multiply\epsfysize\epsftmp   
     \multiply\epsftmp\epsftsize \advance\epsfrsize-\epsftmp
     \epsftmp=\epsfxsize
     \loop \advance\epsfrsize\epsfrsize \divide\epsftmp 2
     \ifnum\epsftmp>0
        \ifnum\epsfrsize<\epsftsize\else
           \advance\epsfrsize-\epsftsize \advance\epsfysize\epsftmp \fi
     \repeat     
   \fi
   \ifepsfverbose\message{#1: width=\the\epsfxsize, height=\the\epsfysize}\fi
   \epsftmp=10\epsfxsize \divide\epsftmp\pspoints
   \newcount\figskipcount
      \message{#1 \the\epsfysize  }
   \vbox to\epsfysize{\vfil\hbox to\epsfxsize{%
      \includegraphics{#1}%
      \hfil}}%
\epsfxsize=0pt\epsfysize=0pt}%

{\catcode`\%=12 \global\let\epsfpercent=
\long\def\epsfaux#1#2:#3\\{\ifx#1\epsfpercent
   \def\testit{#2}\ifx\testit\epsfbblit
      \epsfgrab #3 . . . \\%
      \epsffileokfalse
      \global\epsfbbfoundtrue
   \fi\else\ifx#1\par\else\epsffileokfalse\fi\fi}%
\def\epsfgrab #1 #2 #3 #4 #5\\{%
   \global\def\epsfllx{#1}\ifx\epsfllx\empty
      \epsfgrab #2 #3 #4 #5 .\\\else
   \global\def\epsflly{#2}%
   \global\def\epsfurx{#3}\global\def\epsfury{#4}\fi}%
\def\epsfsize#1#2{\epsfxsize}
\def\figinsert#1#2{\epsfbox{#1} \message{#2} }
\begin{document}

\title
[The ELAIS Deep X-ray Survey I]
{The ELAIS Deep X-ray Survey I: \\
Chandra Source Catalogue and First Results}
\author[J. Manners et al.]
{J.C. Manners$^{1*}$, O. Johnson$^1$, O. Almaini$^1$, C.J. Willott$^2$,
E. Gonzalez-Solares$^3$,
\medskip
\\
\normalfont \LARGE A. Lawrence$^1$, R.G. Mann$^1$,
I. Perez-Fournon$^4$, J.S. Dunlop$^1$, R.G. McMahon$^5$,
\medskip
\\
\normalfont \LARGE S.J. Oliver$^3$, M. Rowan-Robinson$^6$, 
S. Serjeant$^7$
\smallskip
\\
$^1$ Institute for Astronomy, University of Edinburgh, 
Royal Observatory, Blackford Hill, Edinburgh EH9 3HJ
\\
$^2$ Astrophysics, Department of Physics, Keble Rd, Oxford OX1 3RH
\\
$^3$ Astronomy Centre, CPES, University of Sussex, Falmer, Brighton, BN1
9QJ
\\
$^4$ Instituto de Astrofisica de Canarias, 38200 La Laguna, Tenerife, Spain
\\
$^5$ Institute of Astronomy, Madingley Road, Cambridge, CB3 0HA
\\
$^6$ Astrophysics Group, Blackett Laboratory, Imperial College, Prince
Consort Rd, London SW7 2BW
\\
$^7$ Unit for Space Sciences and Astrophysics, School of Physical Sciences,
University of Kent, Cantebury, CT2 7NZ
\\
$^*$ Dipartimento di Astronomia, dell'Universita di Padova, 
Vicolo dell'Osservatorio, 2 - 35122 Padova, Italy as of October 2002}
\date{MNRAS accepted}
\maketitle

\begin{abstract}
We present an analysis of two deep (75 ks) Chandra observations of the
European Large Area ISO Survey (ELAIS) fields N1 and N2 as the first
results from the ELAIS deep X-ray survey. This survey is being conducted
in well studied regions with extensive multi-wavelength coverage. Here we
present the Chandra source catalogues along with an analysis of source
counts, hardness ratios and optical classifications. A total of 233 X-ray
point sources are detected in addition to 2 soft extended sources, which
are found to be associated with galaxy clusters. An over-density of sources
is found in N1 with 30\% more sources than N2, which we attribute to
large-scale structure. A similar variance is seen between other deep
Chandra surveys. The source count statistics reveal an increasing
fraction of hard sources at fainter fluxes. The number of galaxy-like 
counterparts also increases dramatically towards fainter fluxes, consistent 
with the emergence of a large population of obscured sources.
\end{abstract}


\section{Introduction}

The results of recent deep X-ray surveys reveal that almost the entire
X-ray background can be resolved into discrete sources.
The ROSAT Deep Survey (Hasinger {\it et al.} 1998) resolved 70 - 80\% of the
0.5 - 2 keV background at a flux level of 1 $\times 10^{-15}$ erg
s$^{-1}$ cm$^{-2}$. Observations with Chandra and XMM-Newton are now
pushing the detection limits even further. In particular, the
unprecedented resolution of Chandra allows extremely deep observations
that are not limited by source confusion. 
This has been exploited in the Chandra Deep Fields
(North, Brandt {\it et al.} 2001, and South, Giacconi {\it et al.} 2002). 
In the Chandra Deep Field-North 2 Msec of data has been accumulated 
reaching a flux limit of $\sim 1.5 \times 10^{-17}$ erg s$^{-1}$ cm$^{-2}$ 
in the 0.5 - 2 keV band (Barger {\it et al.} 2003).
However, the greatest advances have been at higher energies where
Chandra is now beginning to resolve the 2 - 8 keV background. 

The majority of sources resolved by ROSAT were found to have spectra
that were too steep to account for the flat spectrum of the hard X-ray
background. However towards fainter fluxes a new population emerged in
the ROSAT data with intrinsically harder X-ray spectra (Hasinger {\it
et al.} 1993, Almaini {\it et al.} 1996). Chandra is now uncovering a
large number of hard spectrum sources, and the majority of the 2 - 8
keV background has been resolved. Over the flux range 2 $\times
10^{-16}$ to 10$^{-13}$ erg s$^{-1}$ cm$^{-2}$ the contribution of
resolved sources to the 2 - 8 keV background is 1.1 $\times 10^{-11}$
erg s$^{-1}$ cm$^{-2}$ deg$^{-2}$ (Cowie {\it et al.} 2002). This
translates to $\sim$ 65 - 85 per cent of the background as measured by
Vecchi {\it et al.} (1999, Beppo-Sax) and Ueda {\it et al.} (1999, ASCA) 
respectively.

Early spectroscopic observations are finding a majority of the 
sources with hard X-ray spectra to be type II AGN, indicated by the 
presence of narrow lines (Tozzi {\it et al.} 2001, Barger {\it et al.} 
2001a, Hornschemeier {\it et al.} 2001). Most of these are found 
at $z < 1$. However, a considerable fraction of the hard X-ray sources 
are optically faint, probably due to obscuration, and provide challenging 
targets for spectroscopic identification. Sources identified as type I 
AGN display softer X-ray spectra and are observed to have a higher median 
redshift.

There are still a number of unanswered questions relating to the
properties of the hard X-ray populations at longer wavelengths. AGN
with large X-ray absorbing columns do not always appear as type II AGN
in the optical (e.g. Maiolino {\it et al.} 2001, Willott {\it et al.} 2002). 
The relationship between gas and dust
absorption in AGN remains unclear. It is also uncertain where the
absorbed radiation may be re-radiated. Approximately $\sim$ 7 per
cent of X-ray sources in the Chandra Deep Field North are
sub-millimetre sources (Barger {\it et al.} 2001b), however whether this
is the result of reprocessed nuclear emission or due to a starburst
component, is unknown. Almaini {\it et al.} (2003) find evidence for a strong
angular cross-correlation between the X-ray and sub-millimetre
populations. They suggests there may be an evolutionary sequence in
these galaxies between the major episode of star-formation
(sub-millimetre sources) and the onset of quasar activity (X-ray sources).
To more fully understand the nature of these sources will require in-depth
multi-wavelength studies of the X-ray source population. 

We are conducting a deep X-ray survey with Chandra and XMM in two of the
European Large Area ISO Survey (ELAIS) fields, N1 and N2. These high
latitude fields were chosen for their low cirrus emission, and have a
wealth of multi-wavelength data available. Both
fields have been observed with ISO at 7, 15, 90, and 175 $\mu$m
(Oliver {\it et al.} 2000), with the VLA at 1.4 GHz (Ciliegi {\it et al.} 1999,
Ivison {\it et al.} 2002), and have deep g$'$, r$'$, i$'$, H, and K imaging
(Gonzalez-Solares {\it et al.} 2003).  Region N2 has been mapped with SCUBA to
8 mJy at 850$\mu$m (Fox {\it et al.} 2001, Scott {\it et al.} 2001). As well as
the Chandra observations described here, XMM-Newton observations in region N1 
($5\times 30$ ksec pointings) are awaiting scheduling.

In this paper we present the analysis of the Chandra X-ray data and the
Chandra source catalogue. Paper II (Gonzalez-Solares {\it et al.} 2003) will
present details of the optical identifications.

\section{The X-ray data}

The ELAIS Deep X-ray Survey (EDXS) is being conducted in the northern
ELAIS regions N1 and N2. The Chandra data consists of approximately
75 ks exposures in each field. Region N1 was observed on 3-4 August 
2000 (OBS\_ID
888) and N2 on 2-3 August 2000 (OBS\_ID 887). The nominal aimpoints
were N1: 16:10:20.11 +54:33:22.3, and N2: 16:36:46.99 +41:01:33.7. The
ACIS-I chips were used with the addition of ACIS-S2 and ACIS-S4.

Analysis was carried out on data reprocessed with version
R4CU5UPD14.1 of the pipeline processing software. The data were
reduced using the CIAO software package (version 2.1). Bad pixels and
columns were removed and data were filtered to eliminate high
background times. The latter was achieved through constructing a
lightcurve for background regions and identifying periods of intense
background activity due to solar flares. One obvious flaring period was
identified over the course of the observations resulting in the
removal of 1552 seconds from the data in region N1. More stringent 
conditions for the
removal of high background times were not thought necessary
considering the low level of the quiescent background. After
filtering, exposures in fields N1 and N2 were 71.5 ks and 73.4 ks respectively.

Exposure maps were constructed to account for variations in effective
exposure across an image. This incorporates the positions of
bad pixels, dithering, and vignetting. The effective exposure is
significantly affected by the energy of the source counts. To account
for this, an assumed source spectrum is convolved with the quantum
efficiency of the chip and the effective area of the mirrors. The
resulting map provides an estimate of the effective exposure
(cm$^{-2}$ s$^{-1}$) at each point on the image. For our images, we
used a power-law model spectrum, with photon index $\Gamma = 1.7$.

\section{Source detection}

Sources were detected using a wavelet method, specifically the
{\small WAVDETECT} program (Freeman {\it et al.} 2002) included with the CIAO
software package. The ``Mexican Hat'' wavelet function is used, which
consists of a positive core similar to a canonical PSF, surrounded by
a negative annulus. The overall normalisation is zero.
The zero-crossing point is at a radius of $\sqrt{2}$i, and the
minimum at 2i, where i refers to the scale size in pixels. The correlation
of this wavelet function with an image will reveal sources where
correlation values are larger than a pre-defined threshold.

A $2\times 2$ binned image was used, giving a pixel size of 0.984$''$. 
The threshold
for source detection was set such that the probability of erroneously
identifying a given pixel with a source is 9.5$\times 10^{-7}$. This translates
to a mean detection of 1.0 false sources over the region of the 4 ACIS-I
chips. Wavelet scales were chosen at
i = 2, 2$\sqrt{2}$, 4, 4$\sqrt{2}$, 8, 8$\sqrt{2}$, 16, 16$\sqrt{2}$, and
32 pixels. The minimum scale was chosen to enclose $\sim$90\% encircled
energy of an on-axis PSF. Larger scales can then match the increased
size of off-axis and resolved sources. The algorithm also uses an
exposure map to correct for inconsistencies across the chips.

Sources were detected in 3 bands: 0.5 - 8 keV (full band), 0.5 - 2 keV
(soft band), and 2 - 8 keV (hard band). Below 0.5 keV the quantum
efficiency (including the optical blocking filters) of the front
illuminated chips drops off steeply. A steep rise is also observed in
the background rate due to charged particles. Beyond 8 keV the
effective area of the mirrors is sharply decreasing whilst the
background rate is again beginning to rise.

In order to verify that no sources were missed, we also ran a source
detection on the 0.3 - 10 keV band. All sources found were also
detected in the 0.5 - 8 keV band, and overall, fewer sources were
detected.

\subsection{Sample reliability and detection efficiency}
\label{sec:comp}

For the purposes of source detection, counts flagged as cosmic-ray
afterglows were removed from the image. This procedure is known to
also remove several percent of source photons. Therefore, to obtain
reliable measurements of source flux and extent, a second run of the
{\small WAVDETECT} algorithm was performed on an image where the 
flagged counts were reinstated. Only sources obtained in the original
source detection were used.

{\small WAVDETECT} simulations (Freeman {\it et al.} 2002) suggest a 
mean of 1.0 false sources will be detected over the region of the 4 
ACIS-I chips. We also impose a cut-off at a S/N of 3.0, principally, 
for the purpose of defining a flux limit. This has the effect of removing 
a handful of the least significant sources, further improving the
reliability of the sample.

The detection efficiency of {\small WAVDETECT} is yet to be
definitively determined.  Early simulations have been done by
V. Kashyap (private communication) to determine the probability
of detecting sources of given strengths.  In order to gain an indication
of the number of sources missing from our list of detections, we
have made crude extrapolations to these simulations.  These indicate
that we detect $\sim$ 98.5 -- 99.5 \% of the sources with intrinsic
strengths above the S/N limit of our sample.

\section{The Chandra Source Catalogue}


%
%
\begin{figure}
\centering
\centerline{\epsfxsize=8.5 truecm \figinsert{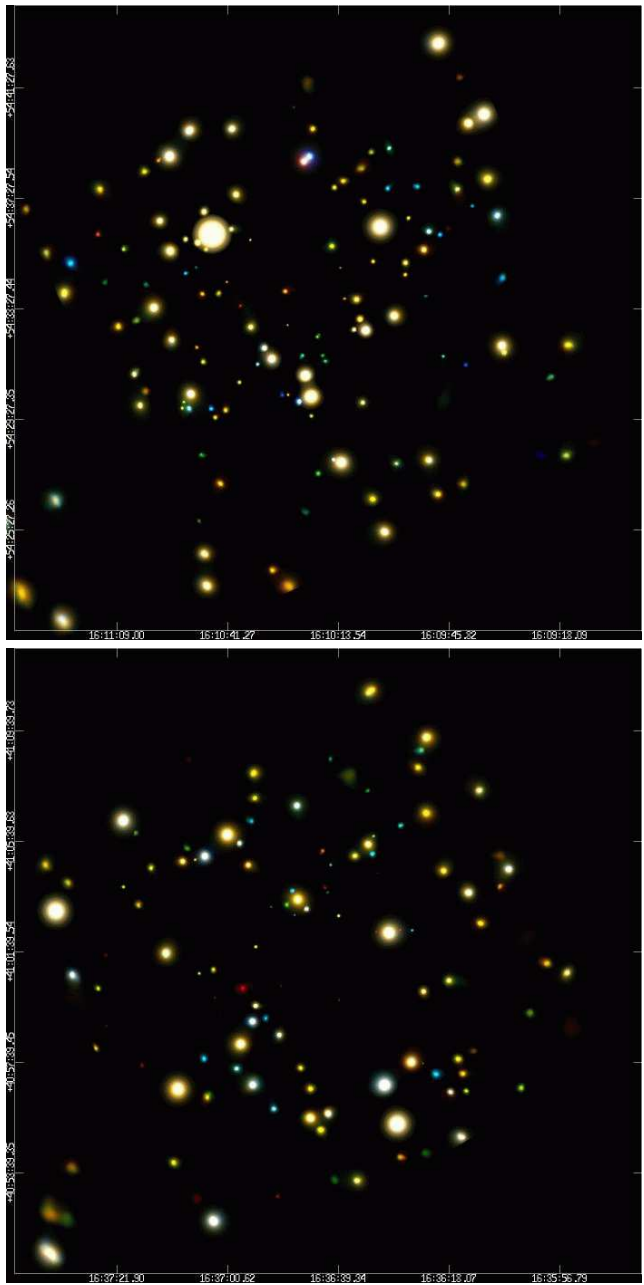}{0.0pt}}
\caption{`True-colour' 
source images of the Chandra fields N1 (top) and N2 (bottom). This is a
noise-free reconstruction of the raw data using the source properties
and correlation images output from the source detection algorithm
{\small WAVDETECT}. 
The colours are constructed from the soft-band (red),
full-band (green), and hard-band (blue).
\label{fig:col}}
\end{figure}

\begin{figure}
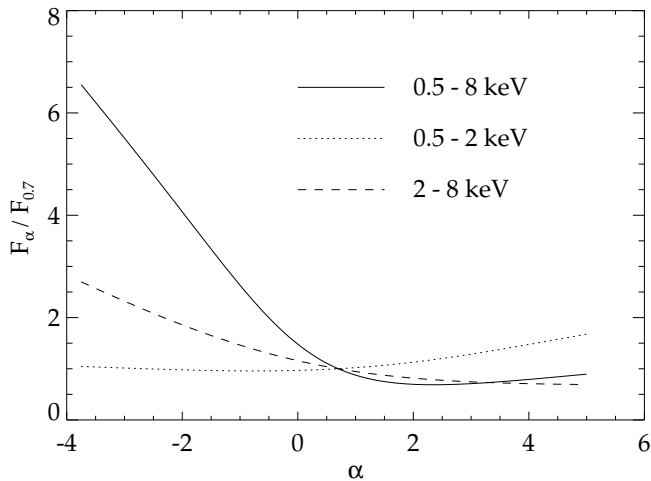

\centering
\centerline{\epsfxsize=8.5 truecm \figinsert{eff.epsi}{0.0pt}}
\caption{Conversion factors to calculate true flux for a source with
spectral slope $\alpha$ given the fluxes quoted in the source
catalogue (which assumes $\alpha = 0.7$). This figure has been 
calculated by passing model spectra of slope $\alpha$ through the total 
response matrix of the detector at a position corresponding to the 
source N1\_23.
\label{fig:fcon}}
\end{figure}

A total of 233 X-ray sources have been detected in the 2 fields 
(Fig.~\ref{fig:col}). In N1
there are 127 sources in the full band (0.5 - 8 keV), 101 in
the soft band (0.5 - 2 keV) including 2 extra sources not
detected in the full band, and 72 sources in the hard band (2
- 8 keV) including 1 extra source not detected in
the full band. There are 57 sources detected in both the soft and hard
bands. In N2 there are 98 sources in the full
band, 81 in the soft band (including 3 extra sources not in the full
band), and 52 sources in the hard band (including  2 extra sources not
in the full band). There are 41 sources detected in both the soft and
hard bands.

The IAU name for the catalogued sources is CXOEN1 JHHMMSS.s+DDMMSS,
for sources in N1 (table 1), and CXOEN2 JHHMMSS.s+DDMMSS for sources
in N2 (table 2). Coordinates are truncated to the above accuracy.

Tables 1 \& 2 display the full catalogue. Sources are detected
to a S/N limit of 3, defined as

\begin{equation}
S/N=C/(1+\sqrt{0.75+B})
\label{eqn:SN}
\end{equation}

\noindent where C are the net source counts, and B the background
counts within the `source cell', a region defined by {\small WAVDETECT} assumed
to contain effectively all of the source counts (Freeman {\it et
al.} 2002). It should be noted that the source cell used here may be
larger than regions used for conventional aperture photometry. The
values calculated for S/N may therefore be lower than those expected
from such methods. The denominator of
equation~\ref{eqn:SN} is an approximate expression for the error on
the background counts (a small number statistic). This comes from
Gehrels (1986): equation (7), which gives the upper confidence level
equivalent to a $1\sigma$ Gaussian error. For sources that do
not reach the S/N limit in a certain band, an upper flux limit has
been calculated from equation 5 (section 5, this paper).

Source coordinates have been astrometrically corrected using
calibrated r$'$ band images (to a depth of r$'$ $\sim$26). High S/N Chandra
sources were matched with stellar r$'$ band counterparts. 16 sources were
used in N1 and 11 in N2, with a good spread across the fields. The
Starlink package ASTROM was used to perform a 6-parameter fit (zero
points, scales in $x$ and $y$, orientation and non-perpendicularity). The RMS
residuals were all less than 1$''$, randomly distributed with a mean 
of $\sim 0.4''$. The
positional error quoted in the catalogue is the error on the centroid
position from the source detection algorithm, with 0.4$''$ added in
quadrature to represent astrometric error.

Net counts are quoted as the total source counts (background
subtracted) in the full energy band (0.5 - 8 keV). Where sources are
only detected in the soft or hard bands, the net counts represent
counts in this band only.

Flux values are calculated assuming a power-law source spectrum of the 
form $F = \nu^{-\alpha}$ with $\alpha = 0.7$ . The effective area will 
vary as a function of $\alpha$ depending on the response of the detector. 
If the slope of the spectrum is known for a given source, Fig.~\ref{fig:fcon} 
can be used to calculate the true flux from the value given in the 
catalogue. This figure has been calculated by passing model spectra of 
slope $\alpha$ through the total response matrix of the detector. For 
illustrative purposes we use the response matrix at a position corresponding 
to the source N1\_23 which lies 4.5 arcmin from the field centre.

\section{Source counts}

In this section we calculate the cumulative source counts, N($>$S), and 
the differential source counts, n(S), in the three bands. We first outline 
details of the calculation
(section~\ref{sec:cal}) which requires knowledge of the available area
of the survey as a function of flux. Section~\ref{sec:lnlsres}
presents our results and comparisons with other surveys, while 
section~\ref{sec:chxrb} describes our source contribution to the hard
X-ray background.

\subsection{Calculating source counts}
\label{sec:cal}

N($>$S) is defined as the sum of the reciprocal areas available for
detecting each source that is brighter than flux S. It follows that
n(S) is the sum of reciprocal areas per flux interval. The sky area
over which a source of flux S may be observed depends on the flux
limit at each point in the image. This, in turn, depends on the
variation in PSF size and effective exposure across the image. 
The flux limit ($S_{lim}$ in erg cm$^{-2}$ s$^{-1}$) may be defined
by the chosen S/N limit of our sample (from equation 1):

\begin{equation}
S/N_{lim}=3=C_{lim}/(1+\sqrt{0.75+B})
\end{equation}

\noindent where

\begin{equation}
C_{lim}=S_{lim}\times\mbox{effective exposure}\times K
\end{equation}

K is a constant (conversion factor from ergs to counts), while the
effective exposure (in cm$^2$s) at each point on the image can be
found from the exposure map. The background counts within the source
region (B), depend on the size of the PSF:

\begin{equation}
B=B_{avg}\times\mbox{PSF size}
\end{equation}

$B_{avg}$ is the mean background counts per pixel. We are left with
the following expression for the flux limit:

\begin{equation}
S_{lim}=3\times\frac{1+\sqrt{0.75+(B_{avg}\times\mbox{PSF size})}}{\mbox{effective exposure}\times K}
\end{equation}

\begin{figure}
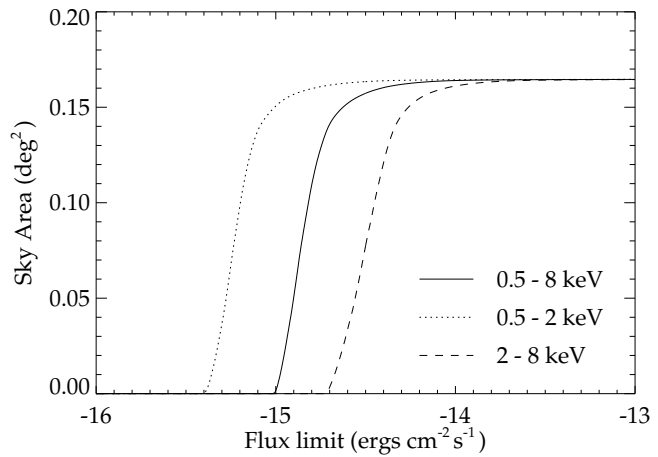

\centering
\centerline{\epsfxsize=8.5 truecm \figinsert{skyarea.epsi}{0.0pt}}
\caption{Sky area observed at survey flux limit.
\label{fig:skya}}
\end{figure}

PSF sizes across the image were taken from the latest PSF library
available with the CIAO software distribution. These were used in
conjunction with the relevant exposure maps for each band to calculate
a `flux limit map' of the Chandra image. The sky area available at a 
given flux limit is found by summing all the pixels with values
smaller than this limit. Fig.~\ref{fig:skya} displays the sky area available at
the flux limit of our survey.

\subsection{Results}
\label{sec:lnlsres}

\begin{figure}
\centering
\centerline{\epsfxsize=8.5 truecm \figinsert{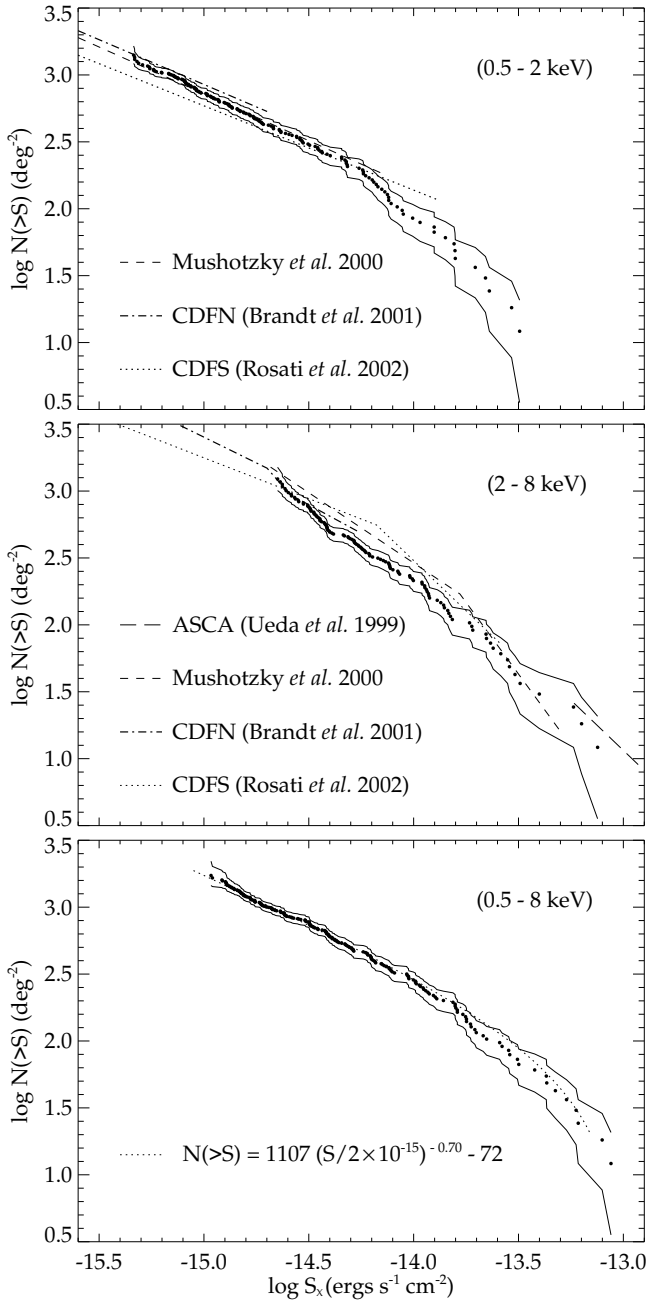}{0.0pt}}
\caption{Cumulative number counts per square degree for sources
detected in the soft (top), hard (middle), and full (bottom) band
images. Data are plotted as filled circles with solid lines enclosing
1$\sigma$ errors. Number counts from other recent surveys are estimated
from best-fitting power laws quoted in the above references. The full 
band plot includes the best-fit line to the differential counts for 
sources with S $< 10^{-13}$ erg s$^{-1}$ cm$^{-2}$.
\label{fig:lnls}}
\end{figure}

The cumulative number counts per square degree are plotted as filled
circles in Fig.~\ref{fig:lnls}. 1$\sigma$ errors are plotted as solid
lines. These incorporate Poisson errors on the counts and the error on
the available sky area. The limiting flux levels are $1.1 \times 10^{-15}$ erg
cm$^{-2}$ s$^{-1}$ (0.5 - 8 keV), $4.6 \times 10^{-16}$ erg
cm$^{-2}$ s$^{-1}$ (0.5 - 2 keV), and $2.2 \times 10^{-15}$ erg
cm$^{-2}$ s$^{-1}$ (2 - 8 keV). Simulations show detection efficiency to be
around 99\% (see section~\ref{sec:comp}), while Eddington bias may result 
in an over-estimation of the cumulative number counts by approximately 
1\% (Manners 2002). These factors work to cancel each other and can safely 
be neglected.

We compare the number counts with those obtained from the Chandra Deep
Field North (CDFN, Brandt {\it et al.} 2001), the Chandra Deep Field South
(CDFS, Rosati {\it et al.} 2002), and those of Mushotzky {\it et al.} 
2000 (M2000).
The number counts in the soft band are in good agreement with CDFN and 
M2000, differing by less than 1$\sigma$ at the flux limit, while CDFS counts 
are $\sim$25\% lower ($\sim$2.5$\sigma$).
The hard band counts of all four surveys are in reasonably good agreement 
at our flux limit. At brighter fluxes the surveys differ by $>2\sigma$, 
most likely as a result of large-scale structure.

\begin{figure}
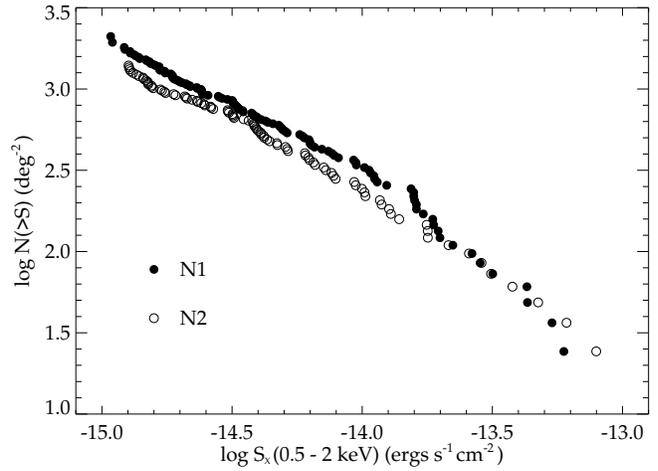

\centering
\centerline{\epsfxsize=8.5 truecm \figinsert{n1_n2_lgnlgs_comp.epsi}{0.0pt}}
\caption{Cumulative source counts (0.5 - 8 keV band) for the
survey fields N1 \& N2 are over-plotted, illustrating the
presence of clustering on scales larger than our field size.
\label{fig:lnlscom}}
\end{figure}

\begin{figure}
\centering
\centerline{\epsfxsize=8.5 truecm \figinsert{n1-n2_hs_comp.epsi}{0.0pt}}
\caption{Comparison of the cumulative source counts for the combined
fields in the soft and hard bands. The counts have been normalised to
an equivalent full band flux (see text, section~\ref{sec:lnlsres}) 
to better emphasise the difference in
slope between these populations.
\label{fig:lnlshs}}
\end{figure}

\begin{figure}
\centering
\centerline{\epsfxsize=8.5 truecm \figinsert{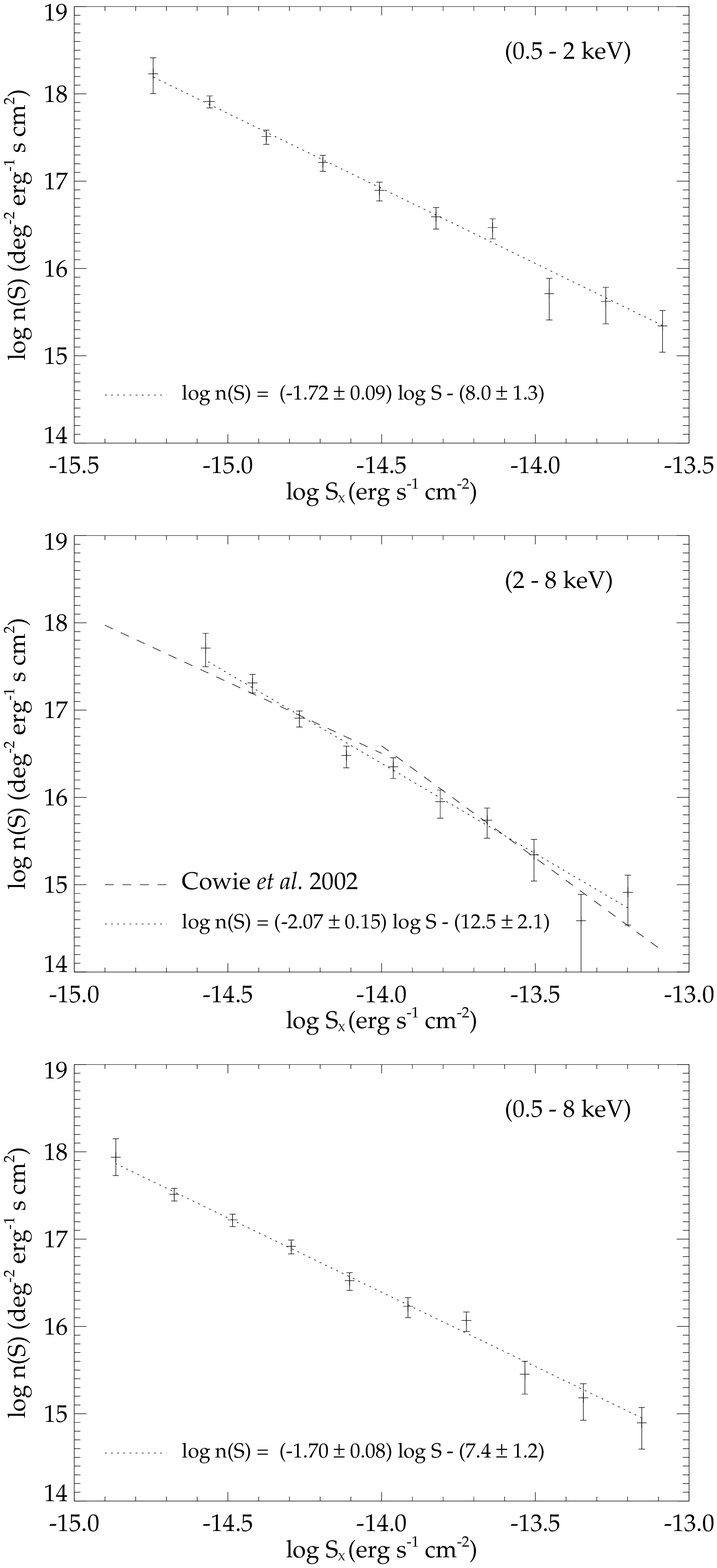}{0.0pt}}
\caption{Differential number counts for sources detected in the soft 
(top), hard(middle), and full band (bottom) images. 1$\sigma$ error bars 
incorporate Poisson errors on the counts and the error on the available 
sky area. A weighted least-squares best fit is displayed with 1$\sigma$ 
error ranges.
\label{fig:difcnts}}
\end{figure}

The presence of clustering on these scales is well illustrated by 
the difference in the number counts observed in N1 and N2
(Fig.~\ref{fig:lnlscom}). 
There are 30\% more sources in N1 than N2 in the full band. In
particular there is an overabundance of brighter sources in N1 at a flux
of $(1-2)\times10^{-14}$ erg cm$^{-2}$ s$^{-1}$. Large-scale structure
is evident in the source images (Fig. 1), most noticeably as a dearth
of sources in the centre of N2. Similar structure can also be seen in images
of the Chandra Deep Field South (Giacconi {\it et al.} 2001). An analysis 
of 9 Chandra fields by Yang {\it et al.} (2003) has shown that such clustering 
is common in the Chandra source population.

A striking feature of the number count relations is the difference in
slope between soft and hard band counts. Fig.~\ref{fig:lnlshs} over-plots 
the soft and hard band counts normalised to an equivalent full band
flux. Normalisation is done in order to plot both populations on the
same flux scale and does not affect the slope of the number
counts. The hard band sources are assumed to have hard spectra and so are
arbitrarily normalised using an alpha of 0. The soft band sources are 
arbitrarily normalised
assuming a soft spectrum with an alpha of 1. The ratio of hard sources
to soft sources is seen to increase dramatically towards fainter fluxes. This
can be explained through the mechanism of obscuration, which will act
to harden the spectra while reducing the flux observed from X-ray sources.

Differential number counts per square degree per unit flux are plotted
in Fig.~\ref{fig:difcnts}. Error bars display 1$\sigma$ errors
incorporating Poisson errors on the counts and the error on the
available sky area. 
The slope of the differential counts for each band was fitted with a
power-law using a weighted least-squares fit. A single power-law was
found to adequately fit the entire flux range for each of the three bands.
For the 0.5 - 2 keV band over the flux range (0.57 - 26) $\times
10^{-15}$ erg cm$^{-2}$ s$^{-1}$ we find:

\begin{equation}
\log {\rm n(S)} = (-1.72 \pm 0.09) \log {\rm S} - (8.0 \pm 1.3)
\end{equation}

For the 2 - 8 keV band over the flux range (2.7 - 63) $\times 10^{-15}$ 
erg cm$^{-2}$ s$^{-1}$ we find:

\begin{equation}
\log {\rm n(S)} = (-2.07 \pm 0.15) \log {\rm S} - (12.5 \pm 2.1)
\end{equation}

For the 0.5 - 8 keV band over the flux range (1.4 - 70) $\times 10^{-15}$ 
erg cm$^{-2}$ s$^{-1}$ we find:

\begin{equation}
\log {\rm n(S)} = (-1.70 \pm 0.08) \log {\rm S} - (7.4 \pm 1.2)
\end{equation}

The hard band differential counts are compared to a maximum likelihood fit 
from Cowie {\it et al.} (2002) to the combined counts from four deep fields 
(CDFN, CDFS, SSA22, \& SSA13). These are found to be in good agreement with 
the error limits of our survey.

\subsection{Contribution to the hard X-ray background}
\label{sec:chxrb}

\begin{figure}
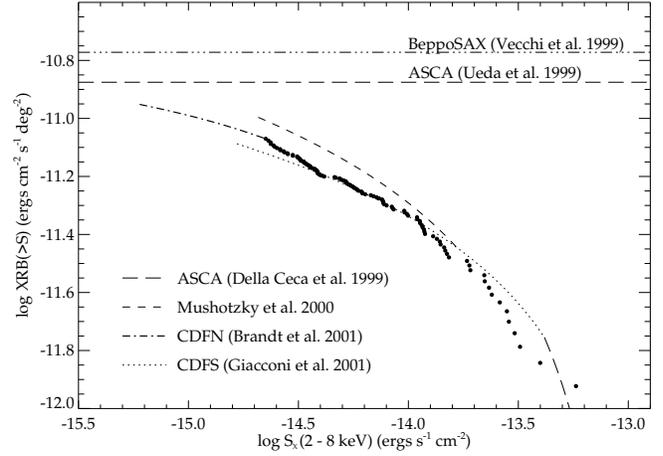

\centering
\centerline{\epsfxsize=8.5 truecm \figinsert{xrb28.epsi}{0.0pt}}
\caption{Contribution to the hard X-ray background. The integrated
hard band flux for sources fainter than $1\times 10^{-13}$ erg
cm$^{-2}$ s$^{-1}$ in this survey are plotted as filled circles. The
contributions from other surveys are extrapolated from reported number
count slopes. Plotted values for the total background exclude sources
brighter than $1\times 10^{-13}$ erg cm$^{-2}$ s$^{-1}$.
\label{fig:xrb}}
\end{figure}

Fig.~\ref{fig:xrb} plots the
integrated source flux for our survey (filled circles) for all sources
with $S<10^{-13}$ erg cm$^{-2}$ s$^{-1}$. At the flux limit of
$S_{2-8} = 2.2 \times 10^{-15}$ erg cm$^{-2}$ s$^{-1}$ the resolved
flux amounts to 8.5 $\times 10^{-12}$ erg cm$^{-2}$ s$^{-1}$
deg$^{-2}$. This is equivalent to between 50 and 64\% of the 2 - 8 keV
background measured by Vecchi {\it et al.} (1999, Beppo-Sax) and Ueda {\it et
al.} (1999, ASCA) respectively. To arrive at these values for the total
background (as plotted in Fig.~\ref{fig:xrb}) the contribution from sources with
$S>10^{-13}$ erg cm$^{-2}$ s$^{-1}$, as observed by ASCA (Della Ceca {\it et
al.} 1999), has been subtracted. The contribution to the background at 
fainter fluxes has been extrapolated from the source counts of the
CDFN survey (Brandt {\it et al.} 2001). By combining the results of our
survey with that of CDFN the contribution to the background within the
flux range $10^{-13}$ - 6 $\times 10^{-16}$ erg cm$^{-2}$ s$^{-1}$
becomes 1.12 $\times 10^{-11}$ erg cm$^{-2}$ s$^{-1}$ deg$^{-2}$,
equivalent to 66\% and 84\% of the aforementioned backgrounds. To
compare the contributions from other surveys, we have used the best
fit to the source counts as published by Mushotzky {\it et al.} (2000), and
Giacconi {\it et al.} (2001, 120ks exposure of CDFS). These have been
normalised at the bright end using the number counts of ASCA sources
from Della Ceca {\it et al.} (1999) to a bright limit of 
$10^{-13}$ erg cm$^{-2}$ s$^{-1}$. The observed discrepancy between the 
different surveys may be due to the effects of clustering on scales larger 
than the survey regions.

\section{Star/galaxy classification}

\begin{figure}
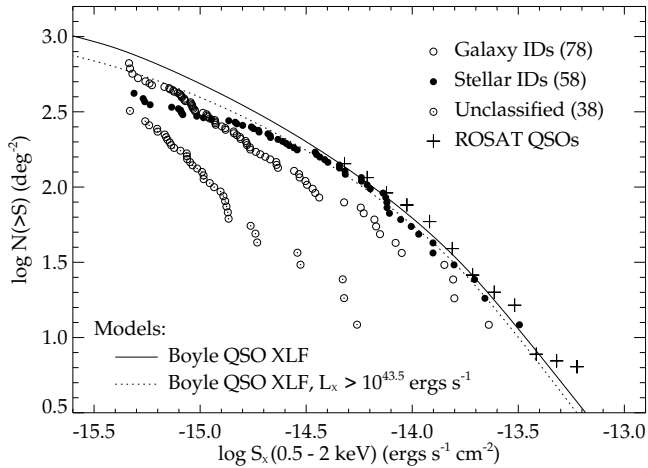

\centering
\centerline{\epsfxsize=8.5 truecm \figinsert{lgnlgs052sg_p.epsi}{0.0pt}}
\caption{Soft band number counts split between
quasar-like, galaxy-like and unclassified (blank fields etc.) optical
counterparts. In comparison the QSOs observed with ROSAT (Boyle {\it et
al.} 1994) and associated models for the QSO X-ray luminosity function
are over-plotted. 
\label{fig:sg}}
\end{figure}

Deep multicolour optical images have been obtained in both our fields,
and will be discussed fully in Gonzalez-Solares {\it et al.} (2003).  They
have identified the optical counterparts of our X-ray sources in r$'$ band
images with limiting magnitudes of $\sim$ 26.  Source optical 
morphologies are classified according to agreement with a stellar point
spread function, as quantified in the SExtractor ``stellarity''
parameter, {\small CLASS STAR} (Bertin \& Arnouts 1996).  The output
of a neural network classifier, the value of this parameter ranges
from 0.0 for significantly extended sources to 1.0 for those with
perfectly stellar PSFs.  For our Chandra sample, we divide sources
with quasar-like and galaxy-like counterparts at {\small CLASS STAR} $=
0.8$.  In practice, morphological classification is increasingly
ambiguous for fainter sources with lower signal to noise and cannot be
considered reliable near to the image limiting magnitude. A number of
X-ray sources remain unclassified where they are associated with
blank fields, gaps in the data, or are near to bright contaminating
sources in the optical images.

The cumulative soft-band source counts for each group have been
calculated and are plotted in Fig.~\ref{fig:sg}.  At bright X-ray
fluxes quasar-like sources are the most numerous. However their number
counts flatten appreciably below a flux of $\sim 5 \times 10^{-15}$
erg cm$^{-2}$ s$^{-1}$. At fainter fluxes the fraction of galaxy-like
sources dramatically increases. At the flux limit of $4.6 \times
10^{-16}$ erg cm$^{-2}$ s$^{-1}$ there are 35\% more galaxy-like
sources than quasar-like sources.

An X-ray luminosity function (XLF) from Boyle {\it et al.} (1994) invoking
pure luminosity evolution, was used to
obtain number count predictions for broad-line AGN. This was based on
observations of 107 QSOs from a deep ROSAT survey. These QSOs reached
a flux limit of $\sim 5 \times 10^{-15}$ erg cm$^{-2}$ s$^{-1}$ in the
0.5 - 2 keV band. We use their best fitting model (model S), with an
exponential decline in the quasar population beyond $z=2.7$, to
construct a prediction for the soft band QSO number counts. This is
over-plotted in Fig.~\ref{fig:sg} (solid line) to compare with the cumulative
number counts for sources with quasar-like optical IDs. 
The Boyle XLF also includes relatively
low luminosity AGN which may possibly be resolved into galaxies in our
r$'$ band images. A second model has therefore been added which excludes
AGN with a 0.5 - 2 keV luminosity less than $10^{43.5}$ erg s$^{-1}$
(dotted line).

The models are well matched to the data up to the flux limit of the
ROSAT survey. However, beyond this they over-predict the number of
quasar-like sources. In the limiting case
where all the unclassified sources are stellar, the data becomes a
good fit to the first model. In the more likely outcome that most of
the unclassified objects are galaxies, the data may still be a
reasonable fit to the Boyle XLF as long as a luminosity cut-off is
applied.

\begin{figure}
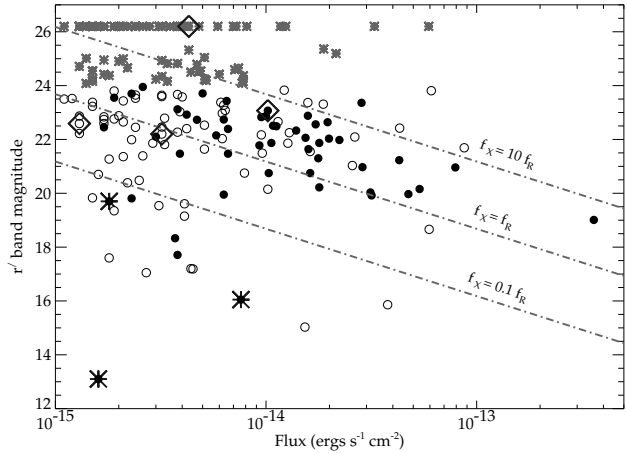

\centerline{\epsfxsize=8.2 truecm \figinsert{fig10.epsi}{0.0pt}}
\caption{X-ray flux vs. r$'$ band magnitude for Chandra sources.
Symbols denote sources with galaxy-like morphology (open circles),
point-like morphology (closed circles), and ambiguous morphology (small
stars), as well as spectroscopically confirmed stars (large stars)
and spectroscopically confirmed Type II AGN (diamonds).
\label{fig:xray_opt}}
\end{figure}

In Fig.~\ref{fig:xray_opt} we plot full band X-ray flux vs. r$'$ band
magnitude for our Chandra sources (compare with Fig. 3 of Barger
{\it et al.} 2002, Fig. 16 of Giacconi {\it et al.} 2002, Fig. 6 of 
Mainieri {\it et al.} 2002). In this figure, as in those in
section \ref{hr}, we indicate source morphology only for sources with
r$'$ $<$ 24, which have unambiguous classifications.  Over-plotted are
lines of constant X-ray to optical ratio, appropriate for the Sloan Gunn r$'$
filter:
\begin{equation}
\log(\frac{f_{X}}{f_{r'}})=\log f_{X} + 5.67 + \frac{r'}{2.5}
\end{equation}
Among point-like
sources optical luminosity is seen to scale with X-ray luminosity;
nearly all exhibit X-ray to optical ratios, $f_{X}/f{r'}$, of 0.1 to
10.    In contrast, sources with galaxy-like
morphology show no tight relation between X-ray and optical fluxes,
suggesting that the host galaxies and not the central AGN dominate the
optical emission.  Four spectroscopically confirmed Type II AGN
(discussed fully in Willott {\it et al.} 2002, Peres-Fournon {\it et al.} 2002)
are marked in Fig. ~\ref{fig:xray_opt}, only one of which has a
notably high value of $f_{X}/f{r'}$.  Of the six point-like sources
with $f_{X}/f{r'} < 0.1$, three are spectroscopically confirmed stars,
as noted in the figure.

\section{Hardness ratios}
\label{hr}
\begin{figure}
\centerline{\psfig{file=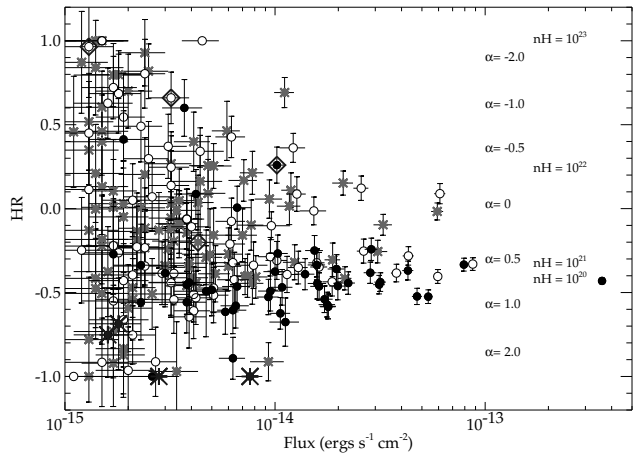,width=9cm,angle=0,clip=}}  
\caption{Full band X-ray flux vs. HR for Chandra sources, with symbols
as in Fig.~\ref{fig:xray_opt} : filled circles are QSOs, open circles
are galaxies, crosses are unclassified sources. 
The values indicated at right are the expected HR of absorbed
power-law spectra at z = 0, with various energy indices assuming
galactic absorption, and with various absorbing
columns assuming an unabsorbed power law of $\alpha = 0.7$.
\label{HR_all}}
\end{figure}

\begin{figure}
\centerline{\psfig{file=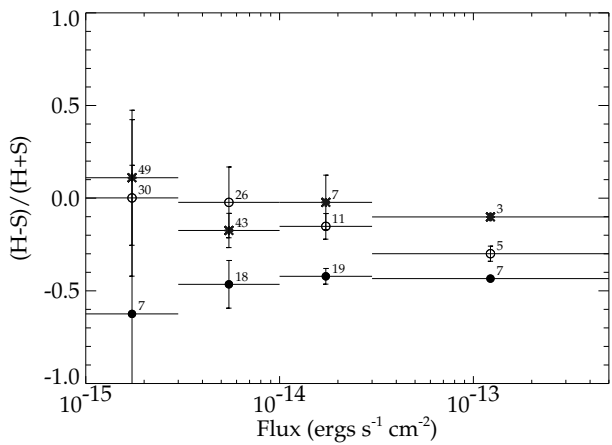,width=9cm,angle=0,clip=}}  
\caption{Error-weighted average hardness ratio for sources of 
different morphology binned by flux, with symbols as in 
Fig.~\ref{HR_all}.  The number of sources in each bin are
indicated, and the error bars indicate the variance within the bin.  
\label{HR_bin}}
\end{figure}

Broad band X-ray hardness ratios were calculated for each source, and
were defined as $ HR = \frac{(H - S)}{(H + S)} $, where H and S are
the background-subtracted source counts in the hard (2.0 - 8.0 keV)
and soft (0.5 - 2.0 keV) bands, respectively.  Net source counts were
extracted from the exposure corrected hard and soft band images within
circular regions centred on the {\small WAVDETECT} positions.  The
apertures were scaled to mimic the degradation of the Chandra PSF with
off-axis angle in each band, and had minimum radii of 10 pixels.
Background counts were extracted from source-free images in annuli
around each source, and subtracted.  The net, source, and background
counts thus obtained were consistent within the errors to those
reported by {\small WAVDETECT}.  We find that background subtraction
and exposure correction of the source counts have increasingly
significant effects on the derived hardness ratios toward fainter
fluxes. 

Hardness ratios are plotted against full band X-ray flux in Fig.~\ref{HR_all}.
As noted by earlier surveys (Mushotzky {\it et al.} 2000,
Giacconi {\it et al.} 2001, Hasinger {\it et al.} 2001, Hornschemeier {\it et
al.} 2001), harder sources are seen at fainter fluxes, signalling the
emergence of the population comprising the majority of the XRB.
Assuming $\alpha = 0.7$ power-law spectra typical of AGN, the range
of hardnesses observed suggests absorbing columns of up to N$_H$ $\sim
10^{23}$ at zero redshift.  As apparent absorption column scales as
$(1+z)^{2.6}$ (see e.g. Barger {\it et al.} 2002), actual columns in higher
redshift sources will be significantly higher.  The range of observed
columns is consistent with that seen in other deep surveys (e.g. Barger 
{\it et al.} 2002, Mainieri {\it et al.} 2002) in which growing
samples of spectroscopic identifications have so far revealed only a
handful of more heavily obscured objects.  

The symbols in Fig.~\ref{HR_all} refer to the morphological
classification discussed in section 5.2. We see that the point-like
sources generally cluster around a HR of -0.5 at all fluxes.  This
value is consistent with a power law of $\alpha$ $\simeq$ 0.7 modified
only by Galactic absorption, and is typical of Type I QSOs.  Sources
with galaxy-like morphology are seen in this region, but also populate
increasingly hard regions of the diagram at fainter fluxes.  Three of
the four confirmed Type II AGN are conspicuously hard.  The trend
to harder X-ray spectra at fainter X-ray fluxes for optically extended
sources is more clearly seen in 
Fig.~\ref{HR_bin}, which shows error-weighted average 
hardness ratios for sources of different morphologies binned by flux. 

\section{Extended Sources}

To search for X-ray sources on scales much larger than the PSF we have
run the source detection algorithm {\small WAVDETECT} using wavelet scales of
16, 16$\sqrt{2}$, 32, 32$\sqrt{2}$, and 64 pixels (see section 3). Any
sources found in addition to those already detected were checked by
inspecting the adaptively smoothed Chandra images.

No additional sources were detected in N1 and inspection of the
smoothed image reveals no hint of large extended sources. In N2 there
are 2 significant extended sources. This is equivalent to $\sim$12 
deg$^{-2}$ over the 2 fields at a limiting soft band flux of 
$\sim 5 \times 10^{-15}$ erg s$^{-1}$ cm$^{-2}$. This source density 
is consistent with the number counts reported by Bauer {\it et al.} 
(2002) and references therein (their Fig. 6).

\begin{figure}
\centering
\centerline{\epsfxsize=8.0 truecm \figinsert{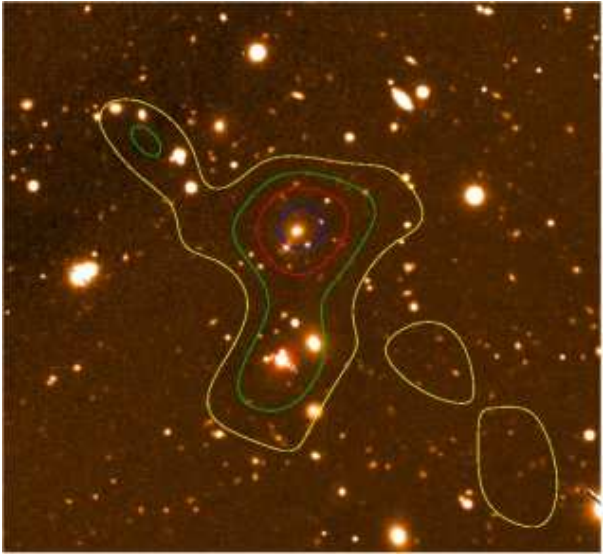}{0.0pt}}
\caption{This 2.5x2.5 arcmin r$'$-band image displays a cluster of
galaxies which coincides with the extended X-ray source CXOEN2
J163637.3+410804. The X-ray contours have been obtained by smoothing
the Chandra data with a Gaussian of 10 arcsec. The X-ray position is
centred on the brightest component of the extended source.
\label{fig:oliv}}
\end{figure}

\begin{figure}
\centerline{\psfig{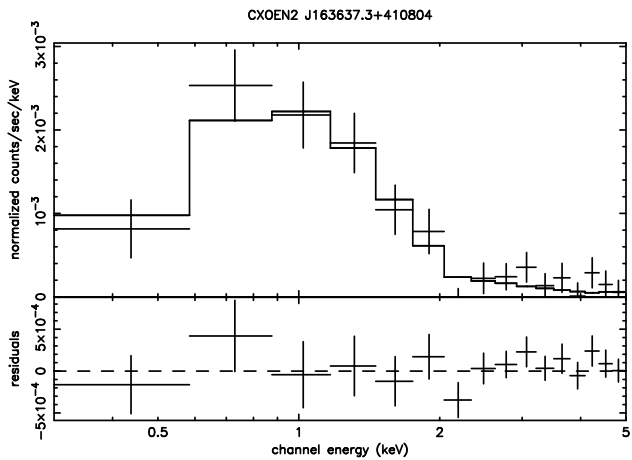}}  
\caption{ACIS-S pulse height spectrum of CXOEN2 J163637.3+410804.  The
model is a Raymond-Smith emission spectrum at the cluster redshift
with 0.3 solar abundance, and yields a best fit plasma temperature of
2.73 keV.   
\label{clus_fit}}
\end{figure}

The most significant of the 2 detected sources is CXOEN2 J163637.3+410804 
displayed in Fig.~\ref{fig:oliv}. The X-ray
position (centred on the brightest component of the extended source)
is at J2000 16:36:37.38 +41:08:04.9. There are 207 net source counts
in the 0.5 - 2 keV band in an area corresponding to a factor 17.5
larger than the PSF. This extrapolates to a soft band flux of $1.46
\pm 0.15 \times 10^{-14}$ erg cm$^{-2}$ s$^{-1}$, although this
includes the flux from point source N2\_101 (CXOEN2 J163633.8+410730)
which lies within the extended source region at a distance of 52
arcsec from the core. The r$'$ band image of this region reveals the
presence of a galaxy cluster.  

Spectra of three cluster members show absorption features consistent
with a redshift of 0.4232 (Perez-Fournon {\it et al.} 2003).
The ACIS-S pulse height spectrum of the cluster
(Fig.~\ref{clus_fit}) and appropriately weighted response matrices
were extracted using standard CIAO tools, and spectral analysis was
performed using XSPEC.  The data is well fitted with a Raymond-Smith
emission model for hot, diffuse gas with an abundance of 0.3 solar and
a plasma temperature of 2.73$\pm$0.81 keV.

\begin{figure}
\centering
\centerline{\epsfxsize=8.0 truecm \figinsert{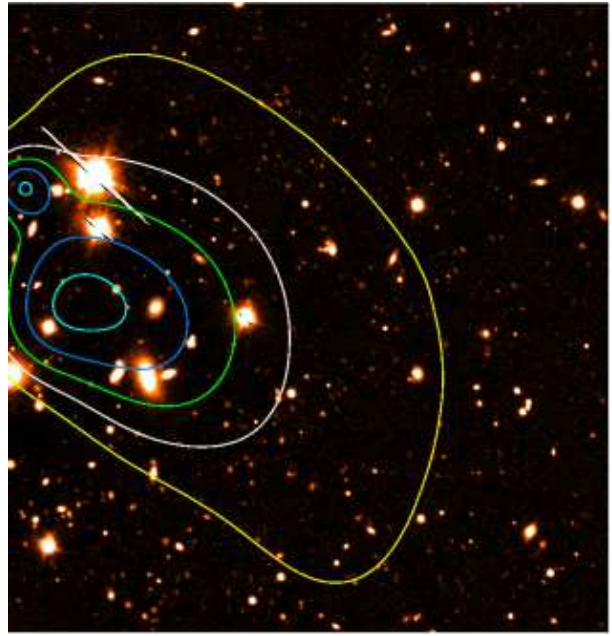}{0.0pt}}
\caption{This 3x3 arcmin r$'$-band image displays a cluster of
galaxies which coincides with an extended X-ray source very close to
the edge of the N2 Chandra image. The X-ray contours have been
obtained from an adaptively smoothed Chandra image and can only be
used as an illustration due to edge effects. The centroid of the X-ray
source is at approximately J2000 16:37:28.5 +41:00:13.
\label{fig:clus2}}
\end{figure}

The second extended X-ray source in N2 is also associated with a
galaxy cluster. This source (shown in Fig.~\ref{fig:clus2}) is at the
very edge of the Chandra image. Its position is approximately J2000
16:37:28.5 +41:00:13, although the centroid may in fact lie outside
the area of the image. For this reason, the identification and
properties of this source will not be reported here.

\section{Conclusions}

We have presented the Chandra source catalogues for deep (75 ks)
observations of the ELAIS fields N1 and N2. A total of 233 X-ray point
sources are detected: 225 in the 0.5 - 8 keV band, 182 in the 0.5 - 2
keV band, and 124 in the 2 - 8 keV band. In addition, 2 extended
sources are detected in N2 in the 0.5 - 2 keV band and are found to be
associated with galaxy clusters.

An over-density of sources is found in N1 with 30\% more sources than
N2. This difference is present in both the soft and hard band number
counts and may be attributed to large-scale structure. A similar 
variance is seen between other deep Chandra surveys.

Source count statistics reveal an increasing fraction of hard sources 
at fainter fluxes. The number of galaxy-like counterparts also increases 
dramatically towards fainter fluxes, consistent with the emergence of a 
large population of obscured sources. Additionally, objects
with galaxy-like and faint optical counterparts exhibit harder X-ray
spectra towards fainter fluxes, consistent with significant absorbing 
columns in this population.

\bigskip

\noindent The source catalogues and further information regarding the
ELAIS deep X-ray survey can be found at this URL:
http://www.roe.ac.uk/$\sim$jcm/edxs

\section*{ACKNOWLEDGEMENTS}
JM acknowledges the support of a PPARC Studentship and would like to
thank the IfA, Edinburgh for providing extra funds for the completion
of this paper.

\clearpage

\begin{sidewaystable*}
\centering
\begin {tabular}{ccccccccccr}
\multicolumn{11}{l}{{\bf Table 1.} Chandra sources in the ELAIS N1 field.} \\
\hline
&  & {\bf RA} & {\bf Dec} & {\bf Err} & {\bf Net} &  & 
\multicolumn{3}{c}{\bf Flux ($\times 10^{-14}$ erg cm$^{-2}$ s$^{-1}$)}  
& \\
{\bf ID} & {\bf CXOEN1} & {\bf (J2000)} & {\bf (J2000)} & 
{\bf (arcsec)} & {\bf Cts} & {\bf S/N} & {\bf (0.5--8keV)}  
& {\bf (0.5--2keV)} & {\bf (2--8keV)} & {\bf HR\hspace{0.6cm}} \\
\hline
\hline
N1\_1 & J161121.8+543402 & 16:11:21.88 & +54:34:02.7 &  0.65 &   76.3 &  17.4 &  1.24 $\pm$ 0.15 &  0.48 $\pm$ 0.07 & $< 0.41$ & $-0.32 \pm 0.10$ \\
N1\_2 & J161113.1+543748 & 16:11:13.10 & +54:37:48.7 &  0.84 &   30.5 &   7.2 &  0.42 $\pm$ 0.09 &  0.12 $\pm$ 0.03 & $< 0.40$ & $-0.44 \pm 0.17$ \\
N1\_3 & J161104.3+543107 & 16:11:04.33 & +54:31:07.2 &  0.63 &   31.9 &   8.8 &  0.43 $\pm$ 0.08 &  0.12 $\pm$ 0.03 &  0.32 $\pm$ 0.12 & $ 0.04 \pm 0.16$ \\ 
N1\_4 & J161059.5+543332 & 16:10:59.53 & +54:33:32.4 &  0.45 &  122.5 &  31.6 &  1.62 $\pm$ 0.15 &  0.59 $\pm$ 0.06 &  1.13 $\pm$ 0.20 & $-0.35 \pm 0.09$ \\ 
N1\_5 & J161058.1+543640 & 16:10:58.16 & +54:36:40.7 &  0.47 &   60.6 &  18.5 &  0.81 $\pm$ 0.11 &  0.31 $\pm$ 0.05 &  0.58 $\pm$ 0.14 & $-0.41 \pm 0.13$ \\ 
N1\_6 & J161055.7+543901 & 16:10:55.74 & +54:39:01.0 &  0.47 &  195.6 &  43.7 &  2.64 $\pm$ 0.19 &  0.89 $\pm$ 0.08 &  2.23 $\pm$ 0.29 & $-0.25 \pm 0.07$ \\ 
N1\_8 & J161055.5+543535 & 16:10:55.50 & +54:35:35.6 &  0.44 &  125.9 &  34.8 &  1.72 $\pm$ 0.16 &  0.70 $\pm$ 0.07 &  0.96 $\pm$ 0.19 & $-0.54 \pm 0.08$ \\ 
N1\_9 & J161055.0+543222 & 16:10:55.09 & +54:32:22.3 &  0.50 &   56.6 &  16.2 &  0.74 $\pm$ 0.10 &  0.28 $\pm$ 0.04 &  0.52 $\pm$ 0.14 & $-0.40 \pm 0.13$ \\ 
N1\_10 & J161052.3+542953 & 16:10:52.37 & +54:29:53.8 &  0.63 &   12.6 &   4.1 &  0.17 $\pm$ 0.05 &  0.05 $\pm$ 0.02 & $< 0.28$ & $-0.27 \pm 0.28$ \\
N1\_11 & J161051.6+543600 & 16:10:51.68 & +54:36:00.9 &  0.57 &   27.0 &   7.9 &  0.35 $\pm$ 0.07 &  0.11 $\pm$ 0.03 &  0.36 $\pm$ 0.12 & $-0.09 \pm 0.20$ \\ 
N1\_12 & J161050.7+542953 & 16:10:50.73 & +54:29:53.9 &  0.61 &   30.8 &   8.5 &  0.41 $\pm$ 0.08 &  0.07 $\pm$ 0.02 &  0.61 $\pm$ 0.15 & $ 0.40 \pm 0.18$ \\ 
N1\_13 & J161050.8+543956 & 16:10:50.85 & +54:39:56.6 &  0.54 &  113.8 &  27.3 &  1.54 $\pm$ 0.15 &  0.58 $\pm$ 0.07 &  1.30 $\pm$ 0.22 & $-0.25 \pm 0.09$ \\ 
N1\_14 & J161050.2+543024 & 16:10:50.21 & +54:30:24.1 &  0.44 &  119.6 &  34.4 &  1.96 $\pm$ 0.18 &  0.73 $\pm$ 0.08 &  1.40 $\pm$ 0.25 & $-0.36 \pm 0.09$ \\ 
N1\_15 & J161048.6+543553 & 16:10:48.64 & +54:35:53.2 &  0.46 &   83.0 &  23.0 &  1.06 $\pm$ 0.12 &  0.45 $\pm$ 0.06 &  0.50 $\pm$ 0.14 & $-0.62 \pm 0.09$ \\ 
N1\_16 & J161047.6+542813 & 16:10:47.68 & +54:28:13.0 &  0.66 &   10.4 &   3.7 &  0.15 $\pm$ 0.05 & $< 0.06$ & $< 0.32$ & $ 0.60 \pm 0.22$ \\
N1\_17 & J161047.5+543401 & 16:10:47.50 & +54:34:01.9 &  0.66 &   12.4 &   4.0 &  0.16 $\pm$ 0.05 &  0.08 $\pm$ 0.02 & $< 0.24$ & $-0.75 \pm 0.25$ \\
N1\_18 & J161047.2+543134 & 16:10:47.25 & +54:31:34.7 &  0.63 &   18.7 &   5.8 &  0.24 $\pm$ 0.06 &  0.07 $\pm$ 0.02 & $< 0.25$ & $-0.23 \pm 0.23$ \\
N1\_19 & J161047.0+543700 & 16:10:47.08 & +54:37:00.8 &  0.52 &   50.9 &  14.1 &  0.66 $\pm$ 0.10 &  0.25 $\pm$ 0.04 &  0.40 $\pm$ 0.12 & $-0.46 \pm 0.13$ \\ 
N1\_20 & J161046.5+543538 & 16:10:46.57 & +54:35:38.8 &  0.69 &   18.2 &   5.4 &  0.23 $\pm$ 0.06 &  0.08 $\pm$ 0.03 & $< 0.24$ & $-0.34 \pm 0.24$ \\
N1\_21 & J161046.0+542328 & 16:10:46.03 & +54:23:28.5 &  0.73 &  126.9 &  18.9 &  1.87 $\pm$ 0.19 &  0.66 $\pm$ 0.08 &  1.10 $\pm$ 0.25 & $-0.44 \pm 0.08$ \\ 
N1\_22 & J161045.1+542952 & 16:10:45.18 & +54:29:52.6 &  0.61 &   18.6 &   5.7 &  0.24 $\pm$ 0.06 & $< 0.05$ &  0.54 $\pm$ 0.14 & $ 0.93 \pm 0.20$ \\
N1\_23 & J161045.1+543612 & 16:10:45.15 & +54:36:12.9 &  0.40 & 2826.1 & 569.9 & 36.00 $\pm$ 0.68 & 14.10 $\pm$ 0.31 & 22.46 $\pm$ 0.83 & $-0.43 \pm 0.02$ \\ 
N1\_24 & J161044.1+542934 & 16:10:44.18 & +54:29:34.1 &  0.70 &   18.9 &   5.5 &  0.24 $\pm$ 0.06 &  0.07 $\pm$ 0.02 & $< 0.26$ & $-0.12 \pm 0.24$ \\
N1\_25 & J161044.1+543601 & 16:10:44.14 & +54:36:01.9 &  0.42 &  148.2 &  41.5 &  1.88 $\pm$ 0.16 &  0.68 $\pm$ 0.07 &  1.45 $\pm$ 0.22 & $-0.31 \pm 0.10$ \\ 
N1\_26 & J161042.8+542710 & 16:10:42.87 & +54:27:10.2 &  0.68 &   24.6 &   7.1 &  0.33 $\pm$ 0.07 &  0.13 $\pm$ 0.03 & $< 0.32$ & $-0.54 \pm 0.20$ \\
N1\_27 & J161041.6+542950 & 16:10:41.62 & +54:29:50.1 &  0.69 &   19.9 &   6.2 &  0.30 $\pm$ 0.07 &  0.09 $\pm$ 0.03 & $< 0.30$ & $-0.34 \pm 0.23$ \\
N1\_28 & J161041.3+543428 & 16:10:41.33 & +54:34:28.4 &  0.55 &    8.8 &   3.2 &  0.11 $\pm$ 0.04 &  0.05 $\pm$ 0.02 & $< 0.22$ & $-1.00 \pm 0.00$ \\
N1\_29 & J161040.2+543623 & 16:10:40.29 & +54:36:23.3 &  0.48 &   25.9 &   8.7 &  0.33 $\pm$ 0.07 &  0.10 $\pm$ 0.03 &  0.36 $\pm$ 0.11 & $-0.19 \pm 0.20$ \\ 
N1\_30 & J161040.1+544000 & 16:10:40.14 & +54:40:00.9 &  0.54 &   70.9 &  18.7 &  0.95 $\pm$ 0.12 &  0.35 $\pm$ 0.05 &  0.77 $\pm$ 0.17 & $-0.29 \pm 0.11$ \\ 
N1\_31 & J161039.1+543738 & 16:10:39.13 & +54:37:38.5 &  0.48 &   61.2 &  17.1 &  0.78 $\pm$ 0.10 &  0.29 $\pm$ 0.05 &  0.49 $\pm$ 0.13 & $-0.45 \pm 0.12$ \\ 
N1\_32 & J161038.1+543050 & 16:10:38.14 & +54:30:50.3 &  0.49 &   15.5 &   5.5 &  0.19 $\pm$ 0.05 &  0.08 $\pm$ 0.02 & $< 0.23$ & $-0.83 \pm 0.27$ \\
N1\_34 & J161035.4+543250 & 16:10:35.40 & +54:32:50.7 &  0.46 &   39.8 &  13.3 &  0.50 $\pm$ 0.08 &  0.19 $\pm$ 0.04 &  0.32 $\pm$ 0.10 & $-0.48 \pm 0.15$ \\ 
\hline
\end{tabular}
\end{sidewaystable*}

\begin{sidewaystable*}
\centering
\begin {tabular}{ccccccccccr}
\multicolumn{11}{l}{{\bf Table 1.} Chandra sources in the ELAIS N1 field (continued).} \\
\hline
&  & {\bf RA} & {\bf Dec} & 
{\bf Err} & {\bf Net} &  & 
\multicolumn{3}{c}{\bf Flux ($\times 10^{-14}$ erg cm$^{-2}$ s$^{-1}$)}  
& \\
{\bf ID} & {\bf CXOEN1} & {\bf (J2000)} & {\bf (J2000)} & 
{\bf (arcsec)} & {\bf Cts} & {\bf S/N} & {\bf (0.5--8keV)}  
& {\bf (0.5--2keV)} & {\bf (2--8keV)} & {\bf HR\hspace{0.6cm}} \\
\hline
N1\_38 & J161033.6+543129 & 16:10:33.67 & +54:31:29.9 &  0.44 &   15.2 &   5.8 &  0.19 $\pm$ 0.05 & $< 0.04$ &  0.32 $\pm$ 0.10 & $ 0.55 \pm 0.26$ \\
N1\_39 & J161031.9+543204 & 16:10:31.97 & +54:32:04.7 &  0.45 &   48.7 &  16.1 &  0.77 $\pm$ 0.11 &  0.23 $\pm$ 0.05 &  0.80 $\pm$ 0.18 & $-0.10 \pm 0.15$ \\ 
N1\_40 & J161030.1+543142 & 16:10:30.12 & +54:31:42.0 &  0.42 &  129.0 &  38.0 &  1.58 $\pm$ 0.14 &  0.57 $\pm$ 0.06 &  1.19 $\pm$ 0.19 & $-0.32 \pm 0.08$ \\ 
N1\_41 & J161027.5+543022 & 16:10:27.59 & +54:30:22.4 &  0.67 &    9.8 &   3.3 &  0.12 $\pm$ 0.04 & $< 0.04$ &  0.25 $\pm$ 0.09 & $ 0.87 \pm 0.30$ \\
N1\_43 & J161026.7+543408 & 16:10:26.78 & +54:34:08.1 &  0.50 &   13.7 &   5.0 &  0.17 $\pm$ 0.05 &  0.08 $\pm$ 0.02 & $< 0.20$ & $-0.92 \pm 0.26$ \\
N1\_45 & J161023.2+543008 & 16:10:23.26 & +54:30:08.7 &  0.50 &   45.3 &  13.6 &  0.59 $\pm$ 0.09 &  0.08 $\pm$ 0.02 &  1.00 $\pm$ 0.18 & $ 0.46 \pm 0.18$ \\ 
N1\_46 & J161022.4+543149 & 16:10:22.45 & +54:31:49.2 &  0.52 &   11.6 &   4.2 &  0.15 $\pm$ 0.05 & $< 0.04$ &  0.24 $\pm$ 0.09 & $ 0.40 \pm 0.28$ \\
N1\_47 & J161022.1+543850 & 16:10:22.13 & +54:38:50.8 &  0.46 &   80.3 &  22.7 &  1.02 $\pm$ 0.12 &  0.36 $\pm$ 0.05 &  0.77 $\pm$ 0.16 & $-0.31 \pm 0.11$ \\ 
N1\_48 & J161021.7+543104 & 16:10:21.76 & +54:31:04.7 &  0.41 &  260.8 &  71.9 &  3.17 $\pm$ 0.20 &  1.25 $\pm$ 0.09 &  1.89 $\pm$ 0.24 & $-0.44 \pm 0.06$ \\ 
N1\_49 & J161020.8+543900 & 16:10:20.88 & +54:39:00.9 &  0.46 &   87.0 &  25.6 &  1.11 $\pm$ 0.12 &  0.10 $\pm$ 0.03 &  2.21 $\pm$ 0.27 & $ 0.69 \pm 0.09$ \\ 
N1\_50 & J161020.3+543020 & 16:10:20.34 & +54:30:20.0 &  0.41 &  434.6 & 112.8 &  5.36 $\pm$ 0.26 &  2.20 $\pm$ 0.12 &  2.81 $\pm$ 0.29 & $-0.52 \pm 0.04$ \\ 
N1\_51 & J161020.2+542937 & 16:10:20.25 & +54:29:37.2 &  0.64 &   13.9 &   4.7 &  0.19 $\pm$ 0.06 & $< 0.05$ & $< 0.24$ & $ 0.03 \pm 0.26$ \\
N1\_52 & J161019.9+544001 & 16:10:19.96 & +54:40:01.8 &  0.68 &   23.1 &   6.8 &  0.32 $\pm$ 0.07 &  0.13 $\pm$ 0.03 & $< 0.31$ & $-0.44 \pm 0.20$ \\
N1\_53 & J161018.8+543229 & 16:10:18.81 & +54:32:29.5 &  0.61 &   10.7 &   3.6 &  0.13 $\pm$ 0.04 & $< 0.04$ & $< 0.19$ & $ 0.52 \pm 0.29$ \\
N1\_54 & J161016.7+543136 & 16:10:16.78 & +54:31:36.9 &  0.58 &   11.0 &   3.8 &  0.13 $\pm$ 0.05 & $< 0.04$ & $< 0.20$ & $ 0.35 \pm 0.26$ \\
N1\_55 & J161015.1+543546 & 16:10:15.12 & +54:35:46.4 &  0.51 &   27.6 &   9.1 &  0.33 $\pm$ 0.07 &  0.13 $\pm$ 0.03 & $< 0.20$ & $-0.39 \pm 0.19$ \\
N1\_56 & J161014.6+542802 & 16:10:14.64 & +54:28:02.2 &  0.63 &   25.0 &   7.4 &  0.32 $\pm$ 0.07 &  0.09 $\pm$ 0.03 &  0.37 $\pm$ 0.12 & $ 0.03 \pm 0.23$ \\ 
N1\_57 & J161014.5+543754 & 16:10:14.57 & +54:37:54.0 &  0.60 &   21.4 &   6.5 &  0.28 $\pm$ 0.07 &  0.09 $\pm$ 0.03 & $< 0.25$ & $-0.13 \pm 0.23$ \\
N1\_58 & J161013.0+543459 & 16:10:13.07 & +54:34:59.6 &  0.72 &    9.0 &   3.1 &  0.11 $\pm$ 0.04 & $< 0.04$ & $< 0.20$ & $ 0.46 \pm 0.33$ \\
N1\_59 & J161012.8+542756 & 16:10:12.80 & +54:27:56.5 &  0.43 &  219.2 &  53.6 &  2.84 $\pm$ 0.20 &  1.07 $\pm$ 0.09 &  1.91 $\pm$ 0.25 & $-0.38 \pm 0.06$ \\ 
N1\_60 & J161012.3+543807 & 16:10:12.33 & +54:38:07.9 &  0.60 &   21.0 &   6.6 &  0.28 $\pm$ 0.07 &  0.14 $\pm$ 0.03 & $< 0.26$ & $-1.00 \pm 0.00$ \\
N1\_61 & J161009.5+543245 & 16:10:09.59 & +54:32:45.6 &  0.58 &   26.3 &   7.9 &  0.32 $\pm$ 0.07 &  0.12 $\pm$ 0.03 &  0.23 $\pm$ 0.09 & $-0.44 \pm 0.21$ \\ 
N1\_62 & J161009.0+543350 & 16:10:09.05 & +54:33:50.9 &  0.42 &   39.8 &  16.6 &  1.12 $\pm$ 0.18 &  0.48 $\pm$ 0.08 &  0.51 $\pm$ 0.20 & $-0.68 \pm 0.14$ \\ 
N1\_64 & J161008.1+543307 & 16:10:08.12 & +54:33:07.7 &  0.44 &   51.7 &  17.2 &  0.63 $\pm$ 0.09 &  0.29 $\pm$ 0.04 &  0.22 $\pm$ 0.09 & $-0.89 \pm 0.12$ \\ 
N1\_65 & J161007.4+543006 & 16:10:07.45 & +54:30:06.8 &  0.54 &   25.2 &   8.7 &  0.48 $\pm$ 0.10 &  0.12 $\pm$ 0.03 &  0.59 $\pm$ 0.18 & $ 0.09 \pm 0.20$ \\ 
N1\_66 & J161007.1+543722 & 16:10:07.16 & +54:37:22.8 &  0.65 &   15.2 &   4.8 &  0.19 $\pm$ 0.05 &  0.09 $\pm$ 0.02 & $< 0.24$ & $-0.87 \pm 0.25$ \\
N1\_67 & J161006.7+543243 & 16:10:06.77 & +54:32:43.3 &  0.41 &  182.5 &  55.5 &  2.23 $\pm$ 0.17 &  0.88 $\pm$ 0.08 &  1.35 $\pm$ 0.20 & $-0.44 \pm 0.07$ \\ 
N1\_68 & J161004.8+543513 & 16:10:04.88 & +54:35:13.3 &  0.61 &   15.6 &   5.1 &  0.20 $\pm$ 0.06 &  0.08 $\pm$ 0.02 & $< 0.22$ & $-0.59 \pm 0.26$ \\
N1\_69 & J161003.1+543628 & 16:10:03.18 & +54:36:28.4 &  0.41 &  481.9 & 116.3 &  5.95 $\pm$ 0.27 &  2.30 $\pm$ 0.12 &  3.98 $\pm$ 0.35 & $-0.40 \pm 0.04$ \\ 
N1\_71 & J161002.0+542525 & 16:10:02.03 & +54:25:25.5 &  0.61 &  107.2 &  21.2 &  1.59 $\pm$ 0.16 &  0.61 $\pm$ 0.07 &  0.84 $\pm$ 0.19 & $-0.45 \pm 0.09$ \\ 
N1\_72 & J161001.2+543752 & 16:10:01.26 & +54:37:52.4 &  0.66 &   16.2 &   5.2 &  0.20 $\pm$ 0.06 & $< 0.05$ &  0.38 $\pm$ 0.12 & $ 0.70 \pm 0.22$ \\
N1\_73 & J161000.8+543918 & 16:10:00.86 & +54:39:18.9 &  0.70 &   22.3 &   6.7 &  0.32 $\pm$ 0.07 &  0.06 $\pm$ 0.02 &  0.36 $\pm$ 0.12 & $ 0.25 \pm 0.21$ \\ 
\hline
\end{tabular}
\end{sidewaystable*}

\begin{sidewaystable*}
\centering
\begin {tabular}{ccccccccccr}
\multicolumn{11}{l}{{\bf Table 1.} Chandra sources in the ELAIS N1 field (continued).} \\
\hline
&  & {\bf RA} & {\bf Dec} & 
{\bf Err} & {\bf Net} &  & 
\multicolumn{3}{c}{\bf Flux ($\times 10^{-14}$ erg cm$^{-2}$ s$^{-1}$)}  
& \\
{\bf ID} & {\bf CXOEN1} & {\bf (J2000)} & {\bf (J2000)} & 
{\bf (arcsec)} & {\bf Cts} & {\bf S/N} & {\bf (0.5--8keV)}  
& {\bf (0.5--2keV)} & {\bf (2--8keV)} & {\bf HR\hspace{0.6cm}} \\
\hline
N1\_75 & J160959.6+543315 & 16:09:59.65 & +54:33:15.2 &  0.43 &  125.4 &  35.7 &  1.62 $\pm$ 0.15 &  0.63 $\pm$ 0.07 &  0.96 $\pm$ 0.18 & $-0.46 \pm 0.08$ \\ 
N1\_76 & J160959.0+542754 & 16:09:59.08 & +54:27:54.1 &  0.63 &   15.5 &   6.2 &  0.49 $\pm$ 0.13 &  0.14 $\pm$ 0.05 &  0.56 $\pm$ 0.22 & $-0.27 \pm 0.27$ \\ 
N1\_77 & J160956.9+543444 & 16:09:56.92 & +54:34:44.5 &  0.62 &   14.3 &   5.6 &  0.34 $\pm$ 0.09 &  0.13 $\pm$ 0.04 & $< 0.37$ & $-0.97 \pm 0.30$ \\
N1\_78 & J160956.7+543510 & 16:09:56.79 & +54:35:10.3 &  0.47 &   26.6 &   8.8 &  0.38 $\pm$ 0.08 &  0.15 $\pm$ 0.03 &  0.26 $\pm$ 0.10 & $-0.46 \pm 0.18$ \\ 
N1\_79 & J160956.0+543647 & 16:09:56.09 & +54:36:47.4 &  0.55 &   14.0 &   4.8 &  0.18 $\pm$ 0.05 &  0.07 $\pm$ 0.02 & $< 0.25$ & $-0.69 \pm 0.27$ \\
N1\_81 & J160952.2+543538 & 16:09:52.23 & +54:35:38.6 &  0.47 &   45.0 &  14.2 &  0.58 $\pm$ 0.09 &  0.24 $\pm$ 0.04 &  0.28 $\pm$ 0.10 & $-0.62 \pm 0.13$ \\ 
N1\_82 & J160951.0+542801 & 16:09:51.05 & +54:28:01.9 &  0.55 &   70.4 &  18.1 &  0.95 $\pm$ 0.12 &  0.35 $\pm$ 0.05 &  0.59 $\pm$ 0.15 & $-0.49 \pm 0.12$ \\ 
N1\_83 & J160951.0+543618 & 16:09:51.02 & +54:36:18.8 &  0.62 &   34.7 &   9.5 &  0.48 $\pm$ 0.09 &  0.10 $\pm$ 0.03 &  0.62 $\pm$ 0.15 & $ 0.26 \pm 0.17$ \\ 
N1\_84 & J160948.6+544307 & 16:09:48.68 & +54:43:07.3 &  0.59 &  304.1 &  39.6 &  4.29 $\pm$ 0.26 &  1.57 $\pm$ 0.11 &  2.86 $\pm$ 0.36 & $-0.37 \pm 0.06$ \\ 
N1\_85 & J160947.4+543147 & 16:09:47.46 & +54:31:47.0 &  0.61 &   13.0 &   4.4 &  0.17 $\pm$ 0.05 & $< 0.05$ & $< 0.25$ & $ 0.01 \pm 0.29$ \\
N1\_86 & J160943.8+543749 & 16:09:43.88 & +54:37:49.3 &  0.71 &   28.8 &   7.6 &  0.38 $\pm$ 0.08 &  0.11 $\pm$ 0.03 &  0.39 $\pm$ 0.12 & $-0.06 \pm 0.18$ \\ 
N1\_87 & J160941.0+544013 & 16:09:41.08 & +54:40:13.1 &  0.58 &  147.1 &  26.0 &  1.99 $\pm$ 0.18 &  0.78 $\pm$ 0.08 &  1.18 $\pm$ 0.22 & $-0.46 \pm 0.08$ \\ 
N1\_89 & J160937.2+544032 & 16:09:37.20 & +54:40:32.6 &  0.51 &  307.8 &  47.6 &  4.31 $\pm$ 0.26 &  1.56 $\pm$ 0.11 &  3.22 $\pm$ 0.37 & $-0.28 \pm 0.05$ \\ 
N1\_90 & J160936.2+543812 & 16:09:36.24 & +54:38:12.5 &  0.58 &   80.9 &  18.2 &  1.08 $\pm$ 0.13 &  0.45 $\pm$ 0.06 &  0.49 $\pm$ 0.15 & $-0.47 \pm 0.11$ \\ 
N1\_92 & J160933.9+543652 & 16:09:33.97 & +54:36:52.3 &  0.58 &   47.1 &  12.5 &  0.62 $\pm$ 0.10 &  0.09 $\pm$ 0.03 &  1.13 $\pm$ 0.21 & $ 0.43 \pm 0.12$ \\ 
N1\_93 & J160932.8+543210 & 16:09:32.83 & +54:32:10.4 &  0.49 &  118.5 &  28.1 &  1.58 $\pm$ 0.15 &  0.58 $\pm$ 0.06 &  1.18 $\pm$ 0.21 & $-0.34 \pm 0.09$ \\ 
N1\_94 & J160932.3+543155 & 16:09:32.31 & +54:31:55.5 &  0.66 &   44.9 &  10.9 &  0.60 $\pm$ 0.10 &  0.21 $\pm$ 0.04 & $< 0.32$ & $-0.27 \pm 0.16$ \\
N1\_95 & J160916.3+543211 & 16:09:16.33 & +54:32:11.1 &  0.79 &   44.7 &   9.7 &  0.63 $\pm$ 0.11 &  0.21 $\pm$ 0.04 & $< 0.43$ & $-0.32 \pm 0.13$ \\
N1\_96 & J161120.3+543508 & 16:11:20.30 & +54:35:08.4 &  0.73 &   32.3 &   7.9 &  0.45 $\pm$ 0.09 & $< 0.07$ &  1.16 $\pm$ 0.22 & $ 1.00 \pm 0.00$ \\
N1\_97 & J161108.3+543250 & 16:11:08.38 & +54:32:50.6 &  0.64 &   29.0 &   8.5 &  0.39 $\pm$ 0.08 &  0.18 $\pm$ 0.04 & $< 0.32$ & $-0.64 \pm 0.19$ \\
N1\_98 & J161107.0+543538 & 16:11:07.08 & +54:35:38.6 &  0.71 &   16.2 &   5.2 &  0.22 $\pm$ 0.06 &  0.07 $\pm$ 0.02 & $< 0.32$ & $-0.14 \pm 0.23$ \\
N1\_99 & J161102.8+542959 & 16:11:02.81 & +54:29:59.5 &  0.68 &   43.4 &  11.1 &  0.70 $\pm$ 0.12 &  0.14 $\pm$ 0.03 &  0.53 $\pm$ 0.16 & $-0.16 \pm 0.16$ \\ 
N1\_100 & J161102.0+543826 & 16:11:02.02 & +54:38:26.7 &  0.76 &   23.3 &   6.6 &  0.32 $\pm$ 0.07 &  0.12 $\pm$ 0.03 & $< 0.36$ & $-0.12 \pm 0.18$ \\
N1\_101 & J161051.9+543006 & 16:10:51.99 & +54:30:06.9 &  0.66 &   12.7 &   4.2 &  0.17 $\pm$ 0.05 & $< 0.05$ & $< 0.28$ & $ 0.11 \pm 0.31$ \\
N1\_102 & J161051.6+543446 & 16:10:51.62 & +54:34:46.9 &  0.69 &   14.3 &   4.6 &  0.18 $\pm$ 0.05 & $< 0.05$ &  0.34 $\pm$ 0.11 & $ 0.69 \pm 0.24$ \\
N1\_103 & J161048.2+542547 & 16:10:48.20 & +54:25:47.9 &  0.87 &   24.8 &   6.0 &  0.35 $\pm$ 0.08 &  0.10 $\pm$ 0.03 & $< 0.40$ & $ 0.05 \pm 0.19$ \\
N1\_104 & J161046.6+542437 & 16:10:46.63 & +54:24:37.5 &  0.72 &   74.6 &  14.5 &  1.14 $\pm$ 0.15 &  0.48 $\pm$ 0.06 &  0.64 $\pm$ 0.19 & $-0.39 \pm 0.11$ \\ 
N1\_106 & J161017.4+543149 & 16:10:17.40 & +54:31:49.2 &  0.72 &   11.5 &   3.8 &  0.14 $\pm$ 0.05 & $< 0.04$ & $< 0.20$ & $ 0.21 \pm 0.29$ \\
N1\_107 & J161011.9+543352 & 16:10:11.98 & +54:33:52.5 &  0.67 &    8.8 &   3.2 &  0.13 $\pm$ 0.05 & $< 0.05$ & $< 0.22$ & $-1.00 \pm 0.00$ \\
N1\_108 & J161004.9+542636 & 16:10:04.99 & +54:26:36.9 &  0.65 &   40.4 &  11.5 &  0.64 $\pm$ 0.11 &  0.23 $\pm$ 0.05 & $< 0.39$ & $-0.42 \pm 0.16$ \\
N1\_110 & J160951.7+543358 & 16:09:51.79 & +54:33:58.1 &  0.75 &   10.8 &   3.6 &  0.14 $\pm$ 0.05 & $< 0.05$ &  0.30 $\pm$ 0.10 & $ 1.00 \pm 0.00$ \\
\hline
\end{tabular}
\end{sidewaystable*}

\begin{sidewaystable*}
\centering
\begin {tabular}{ccccccccccr}
\multicolumn{11}{l}{{\bf Table 1.} Chandra sources in the ELAIS N1 field (continued).} \\
\hline
&  & {\bf RA} & {\bf Dec} & 
{\bf Err} & {\bf Net} &  & 
\multicolumn{3}{c}{\bf Flux ($\times 10^{-14}$ erg cm$^{-2}$ s$^{-1}$)}  
& \\
{\bf ID} & {\bf CXOEN1} & {\bf (J2000)} & {\bf (J2000)} & 
{\bf (arcsec)} & {\bf Cts} & {\bf S/N} & {\bf (0.5--8keV)}  
& {\bf (0.5--2keV)} & {\bf (2--8keV)} & {\bf HR\hspace{0.6cm}} \\
\hline
N1\_111 & J160948.7+542647 & 16:09:48.71 & +54:26:47.1 &  0.61 &   28.6 &   8.8 &  0.40 $\pm$ 0.08 &  0.17 $\pm$ 0.04 & $< 0.37$ & $-0.41 \pm 0.16$ \\
N1\_112 & J160948.2+543611 & 16:09:48.20 & +54:36:11.3 &  0.70 &   17.9 &   5.2 &  0.24 $\pm$ 0.06 & $< 0.05$ &  0.47 $\pm$ 0.14 & $ 0.80 \pm 0.20$ \\
N1\_113 & J161126.3+543528 & 16:11:26.36 & +54:35:28.7 &  1.06 &   32.1 &   6.7 &  0.51 $\pm$ 0.11 &  0.15 $\pm$ 0.04 &  0.46 $\pm$ 0.16 & $-0.16 \pm 0.16$ \\ 
N1\_114 & J161103.8+543303 & 16:11:03.83 & +54:33:03.1 &  0.81 &   15.2 &   4.5 &  0.21 $\pm$ 0.06 & $< 0.06$ & $< 0.31$ & $ 0.05 \pm 0.22$ \\
N1\_115 & J161029.8+542401 & 16:10:29.86 & +54:24:01.6 &  0.88 &   23.3 &   6.0 &  0.38 $\pm$ 0.09 &  0.16 $\pm$ 0.04 & $< 0.49$ & $-0.56 \pm 0.18$ \\
N1\_116 & J161021.6+542608 & 16:10:21.67 & +54:26:08.0 &  0.89 &   11.4 &   3.7 &  0.17 $\pm$ 0.06 & $< 0.06$ & $< 0.34$ & $ 0.80 \pm 0.19$ \\
N1\_117 & J161018.4+542733 & 16:10:18.49 & +54:27:33.4 &  0.77 &   18.5 &   5.3 &  0.24 $\pm$ 0.06 & $< 0.05$ & $< 0.28$ & $ 0.20 \pm 0.22$ \\
N1\_118 & J161005.2+543909 & 16:10:05.23 & +54:39:09.7 &  0.85 &   13.6 &   4.1 &  0.17 $\pm$ 0.05 & $< 0.05$ & $< 0.28$ & $-0.22 \pm 0.26$ \\
N1\_119 & J160953.7+543755 & 16:09:53.78 & +54:37:55.6 &  0.68 &   12.8 &   4.3 &  0.17 $\pm$ 0.05 & $< 0.06$ &  0.33 $\pm$ 0.12 & $ 0.72 \pm 0.19$ \\
N1\_121 & J160932.6+543436 & 16:09:32.63 & +54:34:36.9 &  0.78 &   13.1 &   4.2 &  0.18 $\pm$ 0.06 & $< 0.06$ &  0.56 $\pm$ 0.15 & $ 0.80 \pm 0.14$ \\
N1\_123 & J161137.0+542541 & 16:11:37.00 & +54:25:41.6 &  0.99 &  136.6 &  12.4 &  2.11 $\pm$ 0.24 &  0.56 $\pm$ 0.08 &  2.81 $\pm$ 0.49 & $ 0.15 \pm 0.07$ \\ 
N1\_124 & J161131.4+543706 & 16:11:31.40 & +54:37:06.2 &  1.28 &   18.1 &   3.9 &  0.29 $\pm$ 0.09 &  0.15 $\pm$ 0.04 & $< 0.62$ & $-0.39 \pm 0.19$ \\
N1\_125 & J161123.7+542632 & 16:11:23.71 & +54:26:32.7 &  0.99 &   85.2 &   9.7 &  1.22 $\pm$ 0.17 &  0.29 $\pm$ 0.06 &  1.74 $\pm$ 0.35 & $ 0.36 \pm 0.09$ \\ 
N1\_126 & J161122.0+542217 & 16:11:22.00 & +54:22:17.0 &  1.13 &  214.9 &  12.7 &  3.27 $\pm$ 0.33 &  1.14 $\pm$ 0.12 &  2.66 $\pm$ 0.53 & $-0.10 \pm 0.06$ \\ 
N1\_127 & J161101.6+543030 & 16:11:01.64 & +54:30:30.3 &  0.78 &   12.0 &   4.1 &  0.21 $\pm$ 0.07 &  0.09 $\pm$ 0.03 & $< 0.38$ & $-0.54 \pm 0.24$ \\
N1\_128 & J161043.4+543403 & 16:10:43.44 & +54:34:03.9 &  0.72 &   12.8 &   4.0 &  0.16 $\pm$ 0.05 &  0.07 $\pm$ 0.02 & $< 0.23$ & $-0.38 \pm 0.27$ \\
N1\_129 & J161025.7+542328 & 16:10:25.78 & +54:23:28.4 &  1.01 &   43.2 &   9.0 &  0.93 $\pm$ 0.17 &  0.47 $\pm$ 0.09 & $< 0.60$ & $-0.91 \pm 0.11$ \\
N1\_131 & J160943.6+543849 & 16:09:43.63 & +54:38:49.9 &  0.80 &   14.1 &   4.4 &  0.19 $\pm$ 0.06 &  0.07 $\pm$ 0.02 & $< 0.34$ & $-0.05 \pm 0.18$ \\
N1\_132 & J160920.4+543103 & 16:09:20.49 & +54:31:03.3 &  0.82 &   15.9 &   4.6 &  0.23 $\pm$ 0.07 &  0.06 $\pm$ 0.02 &  0.32 $\pm$ 0.13 & $ 0.00 \pm 0.18$ \\ 
N1\_133 & J160916.9+542811 & 16:09:16.98 & +54:28:11.1 &  0.99 &   33.5 &   6.5 &  0.51 $\pm$ 0.11 &  0.11 $\pm$ 0.03 &  0.68 $\pm$ 0.22 & $ 0.25 \pm 0.13$ \\ 
N1\_135 & J161132.0+542309 & 16:11:32.03 & +54:23:09.1 &  1.30 &  197.5 &  11.2 &  3.08 $\pm$ 0.34 &  1.27 $\pm$ 0.13 & --- & $-0.26 \pm 0.07$ \\
N1\_136 & J161007.7+543834 & 16:10:07.74 & +54:38:34.4 &  0.79 &   16.1 &   4.8 &  0.22 $\pm$ 0.06 &  0.07 $\pm$ 0.03 & $< 0.28$ & $-0.23 \pm 0.24$ \\
N1\_137 & J161101.6+543422 & 16:11:01.64 & +54:34:22.9 &  0.80 &   11.7 &   3.6 &  0.15 $\pm$ 0.05 & $< 0.05$ &  0.31 $\pm$ 0.11 & $ 0.46 \pm 0.26$ \\
N1\_138 & J161109.3+542035 & 16:11:09.31 & +54:20:35.4 &  1.42 &   62.6 &   5.0 &  0.96 $\pm$ 0.22 &  0.40 $\pm$ 0.08 & --- & $-0.10 \pm 0.08$ \\
N1\_139 & J160942.5+542709 & 16:09:42.56 & +54:27:09.1 &  0.97 &   17.4 &   4.7 &  0.24 $\pm$ 0.07 &  0.10 $\pm$ 0.03 & $< 0.38$ & $-0.34 \pm 0.20$ \\
N1\_140 & J161058.4+543852 & 16:10:58.45 & +54:38:52.5 &  0.91 &   14.0 &   4.4 &  0.19 $\pm$ 0.06 &  0.08 $\pm$ 0.03 & $< 0.35$ & $-0.66 \pm 0.28$ \\
N1\_141 & J160943.1+544152 & 16:09:43.19 & +54:41:52.3 &  1.20 &   18.4 &   4.0 &  0.26 $\pm$ 0.08 &  0.11 $\pm$ 0.03 & $< 0.48$ & $-1.00 \pm 0.00$ \\
N1\_142 & J161113.5+543612 & 16:11:13.53 & +54:36:12.6 &  0.99 &   13.7 &   3.7 &  0.20 $\pm$ 0.06 &  0.07 $\pm$ 0.03 & $< 0.39$ & $-0.96 \pm 0.26$ \\
N1\_143 & J160940.1+543713 & 16:09:40.15 & +54:37:13.5 &  0.82 &   11.2 &   3.6 &  0.15 $\pm$ 0.05 & $< 0.06$ &  0.35 $\pm$ 0.12 & $ 1.00 \pm 0.00$ \\
N1\_144 & J160941.8+543127 & 16:09:41.83 & +54:31:27.5 &  0.77 &   10.1 &   3.4 &  0.14 $\pm$ 0.05 & $< 0.06$ &  0.34 $\pm$ 0.12 & $ 1.00 \pm 0.00$ \\
N1\_145 & J161048.8+543205 & 16:10:48.84 & +54:32:05.9 &  0.78 &    9.5 &   3.1 &  0.12 $\pm$ 0.05 & $< 0.05$ & $< 0.25$ & $-0.25 \pm 0.31$ \\
N1\_146 & J160946.1+543624 & 16:09:46.18 & +54:36:24.5 &  0.74 &   12.0 &   3.8 &  0.15 $\pm$ 0.05 &  0.06 $\pm$ 0.02 & $< 0.27$ & $-0.18 \pm 0.26$ \\
N1\_147 & J160909.8+542841 & 16:09:09.89 & +54:28:41.4 &  1.67 &   10.9 &   3.2 & $< 0.31$ &  0.09 $\pm$ 0.03 & $< 0.66$ & $-0.19 \pm 0.17$ \\
N1\_148 & J161037.9+543336 & 16:10:37.92 & +54:33:36.9 &  0.67 &   7.1 &   3.1 & $< 0.11$ &  0.05 $\pm$ 0.02 & $< 0.21$ & $-0.36 \pm 0.31$ \\
N1\_149 & J160923.1+542810 & 16:09:23.15 & +54:28:10.4 &  1.01 &   10.2 &   3.0 & $< 0.20$ & $< 0.08$ & 0.36 $\pm$ 0.14 & $1.00 \pm 0.00$ \\
\hline
\hline
\end{tabular}
\end{sidewaystable*}

\begin{sidewaystable*}
\centering
\begin {tabular}{ccccccccccr}
\multicolumn{11}{l}{{\bf Table 2.} Chandra sources in the ELAIS N2 field.} \\
\hline
&  & {\bf RA} & {\bf Dec} & 
{\bf Err} & {\bf Net} &  & 
\multicolumn{3}{c}{\bf Flux ($\times 10^{-14}$ erg cm$^{-2}$ s$^{-1}$)}  
& \\
{\bf ID} & {\bf CXOEN2} & {\bf (J2000)} & {\bf (J2000)} & 
{\bf (arcsec)} & {\bf Cts} & {\bf S/N} & {\bf (0.5--8keV)}  
& {\bf (0.5--2keV)} & {\bf (2--8keV)} & {\bf HR\hspace{0.6cm}} \\
\hline
\hline
N2\_1 & J163733.4+410309 & 16:37:33.41 & +41:03:09.3 &  0.44 &  638.9 & 101.7 &  8.73 $\pm$ 0.35 &  3.20 $\pm$ 0.15 &  6.32 $\pm$ 0.48 & $-0.33 \pm 0.04$ \\ 
N2\_2 & J163730.2+410049 & 16:37:30.27 & +41:00:49.8 &  0.71 &   52.6 &  11.0 &  0.71 $\pm$ 0.11 &  0.16 $\pm$ 0.04 &  0.77 $\pm$ 0.18 & $ 0.17 \pm 0.12$ \\ 
N2\_4 & J163720.5+410402 & 16:37:20.50 & +41:04:02.2 &  0.59 &   11.6 &   4.2 &  0.17 $\pm$ 0.05 &  0.06 $\pm$ 0.02 & $< 0.32$ & $-0.55 \pm 0.28$ \\
N2\_5 & J163720.5+410626 & 16:37:20.58 & +41:06:26.6 &  0.49 &  216.6 &  45.4 &  2.87 $\pm$ 0.20 &  0.99 $\pm$ 0.08 &  2.33 $\pm$ 0.29 & $-0.24 \pm 0.07$ \\ 
N2\_6 & J163715.2+410443 & 16:37:15.24 & +41:04:43.1 &  0.58 &   18.8 &   6.3 &  0.24 $\pm$ 0.06 &  0.08 $\pm$ 0.02 & $< 0.28$ & $-0.23 \pm 0.22$ \\
N2\_7 & J163712.3+410139 & 16:37:12.38 & +41:01:39.2 &  0.43 &   94.0 &  29.6 &  1.29 $\pm$ 0.14 &  0.48 $\pm$ 0.06 &  0.99 $\pm$ 0.19 & $-0.35 \pm 0.10$ \\ 
N2\_8 & J163712.3+410131 & 16:37:12.36 & +41:01:31.7 &  0.54 &   15.6 &   5.6 &  0.21 $\pm$ 0.06 &  0.08 $\pm$ 0.02 & $< 0.26$ & $-0.40 \pm 0.25$ \\
N2\_9 & J163710.0+405643 & 16:37:10.04 & +40:56:43.2 &  0.43 &  338.9 &  73.8 &  4.74 $\pm$ 0.26 &  1.97 $\pm$ 0.12 &  2.39 $\pm$ 0.30 & $-0.52 \pm 0.05$ \\ 
N2\_10 & J163709.2+410457 & 16:37:09.20 & +41:04:57.5 &  0.52 &   41.0 &  12.1 &  0.51 $\pm$ 0.09 &  0.21 $\pm$ 0.04 &  0.26 $\pm$ 0.10 & $-0.47 \pm 0.15$ \\ 
N2\_11 & J163706.7+410501 & 16:37:06.72 & +41:05:01.7 &  0.77 &   10.1 &   3.5 &  0.13 $\pm$ 0.05 & $< 0.05$ & $< 0.25$ & $-0.78 \pm 0.37$ \\
N2\_12 & J163706.0+410054 & 16:37:06.00 & +41:00:54.6 &  0.57 &   10.4 &   3.7 &  0.13 $\pm$ 0.04 & $< 0.05$ & $< 0.23$ & $ 0.11 \pm 0.33$ \\
N2\_13 & J163705.0+405749 & 16:37:05.03 & +40:57:49.2 &  0.61 &   19.7 &   6.1 &  0.25 $\pm$ 0.06 & $< 0.05$ &  0.57 $\pm$ 0.14 & $ 0.82 \pm 0.16$ \\
N2\_14 & J163704.9+410509 & 16:37:04.94 & +41:05:09.0 &  0.44 &   95.4 &  28.8 &  1.19 $\pm$ 0.12 &  0.29 $\pm$ 0.04 &  1.50 $\pm$ 0.22 & $ 0.11 \pm 0.10$ \\ 
N2\_15 & J163704.4+405625 & 16:37:04.41 & +40:56:25.1 &  0.63 &   15.6 &   5.2 &  0.22 $\pm$ 0.06 &  0.08 $\pm$ 0.03 & $< 0.31$ & $-0.18 \pm 0.23$ \\
N2\_16 & J163703.2+410103 & 16:37:03.27 & +41:01:03.3 &  0.48 &   11.8 &   4.3 &  0.14 $\pm$ 0.05 &  0.06 $\pm$ 0.02 & $< 0.22$ & $-0.48 \pm 0.30$ \\
N2\_17 & J163703.1+405157 & 16:37:03.15 & +40:51:57.0 &  0.66 &  180.1 &  25.0 &  2.57 $\pm$ 0.21 &  0.65 $\pm$ 0.07 &  3.04 $\pm$ 0.37 & $ 0.12 \pm 0.07$ \\ 
N2\_18 & J163700.6+410555 & 16:37:00.64 & +41:05:55.7 &  0.42 &  251.7 &  69.1 &  3.13 $\pm$ 0.20 &  1.25 $\pm$ 0.09 &  1.85 $\pm$ 0.24 & $-0.45 \pm 0.06$ \\ 
N2\_19 & J163658.8+405727 & 16:36:58.82 & +40:57:27.8 &  0.54 &   23.3 &   7.5 &  0.32 $\pm$ 0.07 &  0.09 $\pm$ 0.03 &  0.41 $\pm$ 0.12 & $ 0.14 \pm 0.21$ \\ 
N2\_20 & J163658.3+410537 & 16:36:58.31 & +41:05:37.1 &  0.49 &   31.2 &   9.9 &  0.38 $\pm$ 0.07 &  0.12 $\pm$ 0.03 &  0.41 $\pm$ 0.12 & $-0.06 \pm 0.18$ \\ 
N2\_21 & J163658.0+405821 & 16:36:58.07 & +40:58:21.1 &  0.41 &  175.2 &  52.2 &  2.15 $\pm$ 0.16 &  0.83 $\pm$ 0.07 &  1.39 $\pm$ 0.21 & $-0.42 \pm 0.07$ \\ 
N2\_22 & J163656.6+410449 & 16:36:56.63 & +41:04:49.7 &  0.47 &   34.4 &  11.1 &  0.42 $\pm$ 0.08 &  0.17 $\pm$ 0.03 &  0.24 $\pm$ 0.09 & $-0.53 \pm 0.16$ \\ 
N2\_23 & J163656.0+410625 & 16:36:56.04 & +41:06:25.1 &  0.67 &   11.1 &   3.8 &  0.15 $\pm$ 0.05 & $< 0.05$ &  0.38 $\pm$ 0.12 & $ 1.00 \pm 0.00$ \\
N2\_24 & J163655.7+405652 & 16:36:55.72 & +40:56:52.4 &  0.46 &   97.2 &  25.9 &  1.27 $\pm$ 0.13 &  0.32 $\pm$ 0.05 &  1.53 $\pm$ 0.23 & $ 0.09 \pm 0.10$ \\ 
N2\_25 & J163655.7+405910 & 16:36:55.79 & +40:59:10.5 &  0.43 &   84.7 &  24.5 &  1.02 $\pm$ 0.11 &  0.21 $\pm$ 0.04 &  1.47 $\pm$ 0.21 & $ 0.26 \pm 0.11$ \\ 
N2\_26 & J163655.5+410809 & 16:36:55.56 & +41:08:09.9 &  0.61 &   49.2 &  13.2 &  0.63 $\pm$ 0.10 &  0.28 $\pm$ 0.04 & $< 0.30$ & $-0.61 \pm 0.14$ \\
N2\_27 & J163655.3+410714 & 16:36:55.37 & +41:07:14.7 &  0.64 &   28.2 &   8.4 &  0.39 $\pm$ 0.08 &  0.14 $\pm$ 0.03 & $< 0.29$ & $-0.44 \pm 0.19$ \\
N2\_28 & J163655.2+405944 & 16:36:55.21 & +40:59:44.1 &  0.47 &   32.1 &  11.0 &  0.38 $\pm$ 0.07 &  0.11 $\pm$ 0.03 &  0.38 $\pm$ 0.11 & $-0.14 \pm 0.18$ \\ 
N2\_29 & J163655.1+410152 & 16:36:55.16 & +41:01:52.4 &  0.55 &   14.2 &   4.9 &  0.17 $\pm$ 0.05 &  0.06 $\pm$ 0.02 & $< 0.20$ & $-0.23 \pm 0.28$ \\
N2\_32 & J163653.2+405917 & 16:36:53.26 & +40:59:17.3 &  0.55 &   10.7 &   3.9 &  0.13 $\pm$ 0.04 & $< 0.04$ &  0.28 $\pm$ 0.10 & $ 0.96 \pm 0.30$ \\
N2\_33 & J163651.6+405600 & 16:36:51.69 & +40:56:00.4 &  0.69 &   24.0 &   6.7 &  0.31 $\pm$ 0.07 & $< 0.05$ &  0.52 $\pm$ 0.14 & $ 0.37 \pm 0.19$ \\
N2\_34 & J163650.6+405840 & 16:36:50.63 & +40:58:40.6 &  0.47 &   35.9 &  11.2 &  0.43 $\pm$ 0.08 &  0.14 $\pm$ 0.03 &  0.39 $\pm$ 0.11 & $-0.20 \pm 0.17$ \\ 
N2\_35 & J163649.1+410324 & 16:36:49.18 & +41:03:24.3 &  0.83 &   16.2 &   5.7 &  0.44 $\pm$ 0.12 & $< 0.09$ & $< 0.39$ & $ 0.16 \pm 0.32$ \\
N2\_37 & J163647.3+410659 & 16:36:47.30 & +41:06:59.0 &  0.50 &   60.2 &  16.9 &  0.78 $\pm$ 0.10 &  0.17 $\pm$ 0.04 &  1.10 $\pm$ 0.19 & $ 0.21 \pm 0.13$ \\ 
N2\_38 & J163647.1+410334 & 16:36:47.15 & +41:03:34.8 &  0.42 &  142.6 &  41.1 &  1.79 $\pm$ 0.15 &  0.76 $\pm$ 0.07 &  0.85 $\pm$ 0.17 & $-0.59 \pm 0.07$ \\ 
\hline
\end{tabular}
\end{sidewaystable*}

\begin{sidewaystable*}
\centering
\begin {tabular}{ccccccccccr}
\multicolumn{11}{l}{{\bf Table 2.} Chandra sources in the ELAIS N2 field (continued).} \\
\hline
&  & {\bf RA} & {\bf Dec} & 
{\bf Err} & {\bf Net} &  & 
\multicolumn{3}{c}{\bf Flux ($\times 10^{-14}$ erg cm$^{-2}$ s$^{-1}$)}  
& \\
{\bf ID} & {\bf CXOEN2} & {\bf (J2000)} & {\bf (J2000)} & 
{\bf (arcsec)} & {\bf Cts} & {\bf S/N} & {\bf (0.5--8keV)}  
& {\bf (0.5--2keV)} & {\bf (2--8keV)} & {\bf HR\hspace{0.6cm}} \\
\hline
N2\_39 & J163646.5+405729 & 16:36:46.57 & +40:57:29.1 &  0.53 &   21.4 &   6.9 &  0.26 $\pm$ 0.06 &  0.11 $\pm$ 0.03 & $< 0.23$ & $-0.51 \pm 0.21$ \\
N2\_40 & J163645.5+410313 & 16:36:45.51 & +41:03:13.9 &  0.51 &   25.4 &   8.2 &  0.32 $\pm$ 0.07 &  0.09 $\pm$ 0.03 &  0.33 $\pm$ 0.11 & $-0.06 \pm 0.22$ \\ 
N2\_41 & J163644.7+405540 & 16:36:44.73 & +40:55:40.6 &  0.45 &  139.9 &  35.2 &  1.79 $\pm$ 0.16 &  0.76 $\pm$ 0.07 &  0.69 $\pm$ 0.16 & $-0.58 \pm 0.07$ \\ 
N2\_42 & J163644.6+405643 & 16:36:44.68 & +40:56:43.6 &  0.56 &   32.7 &   9.5 &  0.41 $\pm$ 0.08 &  0.17 $\pm$ 0.03 & $< 0.25$ & $-0.61 \pm 0.17$ \\
N2\_43 & J163642.7+405514 & 16:36:42.71 & +40:55:14.9 &  0.58 &   46.7 &  12.8 &  0.60 $\pm$ 0.09 &  0.23 $\pm$ 0.04 & $< 0.29$ & $-0.34 \pm 0.15$ \\
N2\_44 & J163641.3+405550 & 16:36:41.35 & +40:55:50.2 &  0.48 &   80.9 &  21.3 &  1.03 $\pm$ 0.12 &  0.35 $\pm$ 0.05 &  0.86 $\pm$ 0.17 & $-0.27 \pm 0.11$ \\ 
N2\_46 & J163639.3+410259 & 16:36:39.34 & +41:02:59.5 &  0.55 &   13.2 &   4.6 &  0.15 $\pm$ 0.05 &  0.05 $\pm$ 0.02 & $< 0.19$ & $-0.51 \pm 0.30$ \\
N2\_47 & J163636.2+410509 & 16:36:36.21 & +41:05:09.5 &  0.46 &   31.5 &  10.3 &  0.38 $\pm$ 0.07 &  0.15 $\pm$ 0.03 &  0.27 $\pm$ 0.10 & $-0.45 \pm 0.18$ \\ 
N2\_48 & J163633.6+410534 & 16:36:33.66 & +41:05:34.3 &  0.45 &   76.6 &  22.0 &  0.93 $\pm$ 0.11 &  0.40 $\pm$ 0.05 &  0.52 $\pm$ 0.13 & $-0.53 \pm 0.11$ \\ 
N2\_51 & J163630.5+405651 & 16:36:30.54 & +40:56:51.8 &  0.44 &  291.5 &  72.1 &  6.09 $\pm$ 0.37 &  1.58 $\pm$ 0.14 &  7.54 $\pm$ 0.65 & $ 0.09 \pm 0.06$ \\ 
N2\_52 & J163629.7+410222 & 16:36:29.71 & +41:02:22.7 &  0.41 &  312.1 &  79.2 &  3.78 $\pm$ 0.22 &  1.42 $\pm$ 0.10 &  2.60 $\pm$ 0.28 & $-0.38 \pm 0.05$ \\ 
N2\_54 & J163628.1+405527 & 16:36:28.13 & +40:55:27.5 &  0.42 &  604.8 & 121.2 &  7.92 $\pm$ 0.33 &  2.93 $\pm$ 0.14 &  5.78 $\pm$ 0.44 & $-0.33 \pm 0.04$ \\ 
N2\_55 & J163627.4+410615 & 16:36:27.47 & +41:06:15.4 &  0.73 &   18.3 &   5.5 &  0.23 $\pm$ 0.06 & $< 0.05$ &  0.34 $\pm$ 0.11 & $ 0.49 \pm 0.21$ \\
N2\_56 & J163625.4+405741 & 16:36:25.46 & +40:57:41.4 &  0.46 &  138.3 &  34.0 &  1.77 $\pm$ 0.16 &  0.75 $\pm$ 0.07 &  0.76 $\pm$ 0.16 & $-0.57 \pm 0.07$ \\ 
N2\_57 & J163623.0+410015 & 16:36:23.07 & +41:00:15.0 &  0.52 &   32.8 &  10.2 &  0.41 $\pm$ 0.08 &  0.17 $\pm$ 0.03 &  0.25 $\pm$ 0.09 & $-0.51 \pm 0.16$ \\ 
N2\_58 & J163622.5+410641 & 16:36:22.54 & +41:06:41.6 &  0.56 &   59.6 &  14.8 &  0.76 $\pm$ 0.11 &  0.39 $\pm$ 0.05 & $< 0.30$ & $-1.00 \pm 0.00$ \\
N2\_59 & J163622.4+410927 & 16:36:22.49 & +41:09:27.7 &  0.61 &  104.7 &  20.0 &  1.39 $\pm$ 0.15 &  0.55 $\pm$ 0.06 &  0.74 $\pm$ 0.19 & $-0.39 \pm 0.09$ \\ 
N2\_60 & J163619.2+410436 & 16:36:19.23 & +41:04:36.9 &  0.52 &   51.0 &  14.2 &  0.65 $\pm$ 0.10 &  0.26 $\pm$ 0.04 & $< 0.28$ & $-0.58 \pm 0.13$ \\
N2\_61 & J163618.2+410038 & 16:36:18.23 & +41:00:38.6 &  0.57 &   38.5 &  11.0 &  0.52 $\pm$ 0.09 &  0.18 $\pm$ 0.04 &  0.39 $\pm$ 0.12 & $-0.39 \pm 0.17$ \\ 
N2\_62 & J163616.4+405748 & 16:36:16.42 & +40:57:48.3 &  0.59 &   29.9 &   8.8 &  0.39 $\pm$ 0.08 &  0.16 $\pm$ 0.03 & $< 0.31$ & $-0.35 \pm 0.17$ \\
N2\_63 & J163615.6+405716 & 16:36:15.60 & +40:57:16.6 &  0.71 &   20.3 &   6.1 &  0.27 $\pm$ 0.07 &  0.09 $\pm$ 0.03 & $< 0.32$ & $-0.91 \pm 0.21$ \\
N2\_64 & J163614.4+410349 & 16:36:14.46 & +41:03:49.1 &  0.53 &   78.0 &  19.0 &  1.00 $\pm$ 0.12 &  0.37 $\pm$ 0.05 &  0.72 $\pm$ 0.16 & $-0.38 \pm 0.11$ \\ 
N2\_65 & J163612.1+410242 & 16:36:12.16 & +41:02:42.7 &  0.62 &   39.5 &  10.7 &  0.51 $\pm$ 0.09 &  0.20 $\pm$ 0.04 & $< 0.30$ & $-0.52 \pm 0.16$ \\
N2\_66 & J163606.7+410440 & 16:36:06.79 & +41:04:40.0 &  0.71 &   79.4 &  15.0 &  1.17 $\pm$ 0.15 &  0.34 $\pm$ 0.05 &  1.15 $\pm$ 0.23 & $ 0.01 \pm 0.11$ \\ 
N2\_67 & J163555.7+410054 & 16:35:55.72 & +41:00:54.5 &  0.95 &   53.1 &   8.9 &  0.79 $\pm$ 0.13 &  0.23 $\pm$ 0.05 &  0.53 $\pm$ 0.17 & $-0.34 \pm 0.14$ \\ 
N2\_68 & J163725.2+410021 & 16:37:25.26 & +41:00:21.1 &  0.69 &   17.5 &   5.3 &  0.23 $\pm$ 0.06 &  0.08 $\pm$ 0.03 & $< 0.33$ & $-0.56 \pm 0.22$ \\
N2\_71 & J163710.8+405402 & 16:37:10.81 & +40:54:02.4 &  0.77 &   30.7 &   7.1 &  0.42 $\pm$ 0.09 &  0.13 $\pm$ 0.03 & $< 0.40$ & $ 0.09 \pm 0.15$ \\
N2\_72 & J163657.7+410021 & 16:36:57.74 & +41:00:21.5 &  0.73 &   11.9 &   3.7 &  0.15 $\pm$ 0.05 &  0.08 $\pm$ 0.02 & $< 0.21$ & $-0.92 \pm 0.26$ \\
N2\_73 & J163635.8+405325 & 16:36:35.86 & +40:53:25.6 &  0.78 &   46.1 &   9.8 &  0.62 $\pm$ 0.10 &  0.19 $\pm$ 0.04 &  0.35 $\pm$ 0.13 & $-0.08 \pm 0.14$ \\ 
N2\_74 & J163632.9+411111 & 16:36:32.95 & +41:11:11.4 &  0.88 &   53.2 &  10.0 &  0.72 $\pm$ 0.11 &  0.29 $\pm$ 0.05 & $< 0.46$ & $-0.40 \pm 0.11$ \\
N2\_75 & J163632.7+410513 & 16:36:32.78 & +41:05:13.7 &  0.69 &   20.4 &   6.2 &  0.25 $\pm$ 0.06 & $< 0.05$ &  0.32 $\pm$ 0.11 & $ 0.30 \pm 0.22$ \\
\hline
\end{tabular}
\end{sidewaystable*}

\begin{sidewaystable*}
\centering
\begin {tabular}{ccccccccccr}
\multicolumn{11}{l}{{\bf Table 2.} Chandra sources in the ELAIS N2 field (continued).} \\
\hline
&  & {\bf RA} & {\bf Dec} & 
{\bf Err} & {\bf Net} &  & 
\multicolumn{3}{c}{\bf Flux ($\times 10^{-14}$ erg cm$^{-2}$ s$^{-1}$)}  
& \\
{\bf ID} & {\bf CXOEN2} & {\bf (J2000)} & {\bf (J2000)} & 
{\bf (arcsec)} & {\bf Cts} & {\bf S/N} & {\bf (0.5--8keV)}  
& {\bf (0.5--2keV)} & {\bf (2--8keV)} & {\bf HR\hspace{0.6cm}} \\
\hline
N2\_76 & J163632.6+410552 & 16:36:32.64 & +41:05:52.7 &  0.65 &   12.7 &   4.2 &  0.15 $\pm$ 0.05 &  0.05 $\pm$ 0.02 & $< 0.25$ & $-0.22 \pm 0.30$ \\
N2\_77 & J163625.2+410228 & 16:36:25.25 & +41:02:28.2 &  0.68 &   12.3 &   3.9 &  0.15 $\pm$ 0.05 & $< 0.05$ &  0.31 $\pm$ 0.11 & $ 1.00 \pm 0.00$ \\
N2\_78 & J163624.1+410821 & 16:36:24.12 & +41:08:21.2 &  0.77 &   36.3 &   8.3 &  0.47 $\pm$ 0.09 &  0.17 $\pm$ 0.04 & $< 0.35$ & $-0.29 \pm 0.16$ \\
N2\_79 & J163620.6+405714 & 16:36:20.68 & +40:57:14.8 &  0.75 &   24.7 &   7.1 &  0.32 $\pm$ 0.07 & $< 0.06$ &  0.55 $\pm$ 0.14 & $ 0.66 \pm 0.15$ \\
N2\_80 & J163617.9+405636 & 16:36:17.96 & +40:56:36.5 &  0.72 &   30.1 &   8.0 &  0.40 $\pm$ 0.08 &  0.08 $\pm$ 0.03 &  0.39 $\pm$ 0.12 & $-0.11 \pm 0.18$ \\ 
N2\_81 & J163731.1+410410 & 16:37:31.15 & +41:04:10.8 &  0.87 &   21.7 &   5.4 &  0.31 $\pm$ 0.08 &  0.09 $\pm$ 0.03 & $< 0.41$ & $-0.24 \pm 0.18$ \\
N2\_82 & J163718.1+410600 & 16:37:18.16 & +41:06:00.1 &  0.72 &   11.0 &   3.7 &  0.14 $\pm$ 0.05 &  0.05 $\pm$ 0.02 & $< 0.32$ & $-0.00 \pm 0.22$ \\
N2\_83 & J163717.6+410324 & 16:37:17.66 & +41:03:24.1 &  0.83 &   11.7 &   3.9 &  0.17 $\pm$ 0.06 &  0.09 $\pm$ 0.03 & $< 0.32$ & $-0.73 \pm 0.28$ \\
N2\_84 & J163708.3+410526 & 16:37:08.37 & +41:05:26.0 &  0.72 &   11.2 &   3.9 &  0.15 $\pm$ 0.05 & $< 0.05$ & $< 0.27$ & $ 0.13 \pm 0.29$ \\
N2\_85 & J163648.0+410354 & 16:36:48.04 & +41:03:54.9 &  0.71 &   10.8 &   3.6 &  0.14 $\pm$ 0.05 & $< 0.04$ &  0.27 $\pm$ 0.10 & $ 0.84 \pm 0.28$ \\
N2\_86 & J163633.7+411102 & 16:36:33.71 & +41:11:02.4 &  0.83 &   42.6 &   9.0 &  0.61 $\pm$ 0.11 &  0.19 $\pm$ 0.04 &  0.76 $\pm$ 0.20 & $-0.21 \pm 0.12$ \\ 
N2\_87 & J163627.6+405416 & 16:36:27.62 & +40:54:16.1 &  0.99 &   20.8 &   5.0 &  0.30 $\pm$ 0.08 &  0.10 $\pm$ 0.03 & $< 0.37$ & $-0.39 \pm 0.20$ \\
N2\_88 & J163616.0+405500 & 16:36:16.01 & +40:55:00.9 &  0.82 &   57.2 &  11.9 &  0.94 $\pm$ 0.14 &  0.30 $\pm$ 0.06 &  0.62 $\pm$ 0.18 & $ 0.06 \pm 0.11$ \\ 
N2\_89 & J163615.0+405639 & 16:36:15.03 & +40:56:39.3 &  0.81 &   18.5 &   5.3 &  0.26 $\pm$ 0.07 &  0.07 $\pm$ 0.02 & $< 0.36$ & $ 0.07 \pm 0.20$ \\
N2\_90 & J163612.3+410731 & 16:36:12.34 & +41:07:31.3 &  0.81 &   48.7 &   9.5 &  0.66 $\pm$ 0.11 &  0.18 $\pm$ 0.04 &  0.38 $\pm$ 0.13 & $ 0.01 \pm 0.13$ \\ 
N2\_91 & J163604.3+405646 & 16:36:04.31 & +40:56:46.8 &  0.99 &   22.9 &   5.5 &  0.32 $\pm$ 0.08 &  0.09 $\pm$ 0.03 & $< 0.43$ & $ 0.27 \pm 0.17$ \\
N2\_92 & J163559.3+410116 & 16:35:59.30 & +41:01:16.0 &  0.73 &   14.7 &   4.3 &  0.21 $\pm$ 0.06 &  0.09 $\pm$ 0.03 & $< 0.43$ & $-0.75 \pm 0.17$ \\
N2\_93 & J163735.5+410448 & 16:37:35.54 & +41:04:48.9 &  0.97 &   28.4 &   6.5 &  0.39 $\pm$ 0.09 &  0.13 $\pm$ 0.03 & $< 0.45$ & $-0.33 \pm 0.15$ \\
N2\_94 & J163734.5+405046 & 16:37:34.51 & +40:50:46.7 &  0.95 &  395.9 &  22.4 &  5.92 $\pm$ 0.39 &  1.78 $\pm$ 0.14 &  5.54 $\pm$ 0.67 & $-0.02 \pm 0.05$ \\ 
N2\_96 & J163608.4+410404 & 16:36:08.45 & +41:04:04.1 &  0.73 &   13.3 &   4.2 &  0.19 $\pm$ 0.06 &  0.08 $\pm$ 0.02 & $< 0.36$ & $-0.43 \pm 0.22$ \\
N2\_97 & J163602.6+405927 & 16:36:02.66 & +40:59:27.3 &  1.03 &   26.0 &   5.7 &  0.35 $\pm$ 0.08 & $< 0.07$ & $< 0.39$ & $-0.09 \pm 0.18$ \\
N2\_98 & J163623.4+410859 & 16:36:23.47 & +41:08:59.0 &  0.88 &   27.9 &   6.3 &  0.37 $\pm$ 0.08 &  0.06 $\pm$ 0.02 &  0.40 $\pm$ 0.13 & $ 0.60 \pm 0.17$ \\ 
N2\_99 & J163734.5+405212 & 16:37:34.56 & +40:52:12.8 &  1.12 &  108.3 &   8.5 &  1.53 $\pm$ 0.22 &  0.59 $\pm$ 0.08 & --- & $-0.01 \pm 0.08$ \\
N2\_100 & J163729.9+405349 & 16:37:29.99 & +40:53:49.3 &  1.04 &   23.9 &   4.1 &  0.34 $\pm$ 0.10 &  0.20 $\pm$ 0.05 & --- & $-0.01 \pm 0.12$ \\
N2\_101 & J163633.8+410731 & 16:36:33.84 & +41:07:31.0 &  0.86 &   15.4 &   4.7 &  0.21 $\pm$ 0.06 & $< 0.06$ & $< 0.30$ & $-0.47 \pm 0.26$ \\
N2\_102 & J163725.6+405811 & 16:37:25.68 & +40:58:11.0 &  0.83 &   12.4 &   4.9 &  0.47 $\pm$ 0.14 &  0.19 $\pm$ 0.06 & $< 0.69$ & $-0.49 \pm 0.27$ \\
N2\_103 & J163639.9+405322 & 16:36:39.97 & +40:53:22.8 &  0.98 &   14.1 &   4.2 &  0.19 $\pm$ 0.06 & $< 0.06$ & $< 0.38$ & $ 0.41 \pm 0.17$ \\
N2\_104 & J163640.7+410449 & 16:36:40.77 & +41:04:49.8 &  0.76 &   10.6 &   3.5 &  0.13 $\pm$ 0.05 & $< 0.05$ & $< 0.22$ & $ 0.45 \pm 0.31$ \\
N2\_105 & J163640.9+410840 & 16:36:40.90 & +41:08:40.7 &  0.71 &   12.3 &   4.0 &  0.16 $\pm$ 0.05 & $< 0.06$ & $< 0.31$ & $ 0.63 \pm 0.21$ \\
N2\_106 & J163642.3+410520 & 16:36:42.33 & +41:05:20.8 &  0.75 &   11.1 &   3.5 &  0.14 $\pm$ 0.05 &  0.05 $\pm$ 0.02 & $< 0.23$ & $-0.42 \pm 0.29$ \\
N2\_107 & J163608.4+410507 & 16:36:08.41 & +41:05:07.0 &  0.86 &   12.1 &   3.4 &  0.18 $\pm$ 0.06 &  0.10 $\pm$ 0.03 & $< 0.39$ & $-0.26 \pm 0.22$ \\
N2\_108 & J163613.4+405806 & 16:36:13.45 & +40:58:06.4 &  0.82 &    9.7 &   3.2 &  0.13 $\pm$ 0.05 & $< 0.06$ & $< 0.32$ & $-0.13 \pm 0.25$ \\
N2\_109 & J163730.7+405152 & 16:37:30.77 & +40:51:52.6 &  1.23 &   31.2 &   3.9 &  0.44 $\pm$ 0.13 & --- & --- & $ 0.34 \pm 0.09$ \\
N2\_110 & J163708.0+410840 & 16:37:08.04 & +41:08:40.1 &  1.31 & 8.3 &   3.2 &  $< 0.20$ & 0.08 $\pm$ 0.03 & $< 0.48$ & $-0.40 \pm 0.28$ \\
N2\_111 & J163627.5+410228 & 16:36:27.50 & +41:02:28.6 &  0.59 & 7.1 &   3.2 &  $< 0.11$ & 0.05 $\pm$ 0.02 & $< 0.23$ & $-1.00 \pm 0.00$ \\
N2\_112 & J163723.8+410133 & 16:37:23.83 & +41:01:33.3 &  0.88 & 7.6 &   3.1 &  $< 0.14$ & 0.06 $\pm$ 0.02 & $< 0.32$ & $-0.41 \pm 0.26$ \\
N2\_113 & J163621.4+410049 & 16:36:21.42 & +41:00:49.9 &  0.98 & 10.0 &  3.4 &  $< 0.12$ & $< 0.05$ & 0.30 $\pm$ 0.11 & $1.00 \pm 0.00$ \\
N2\_114 & J163631.8+410432 & 16:36:31.82 & +41:04:32.7 &  0.59 & 7.6 &   3.1 &  $< 0.12$ & $< 0.05$ & 0.24 $\pm$ 0.09 & $0.52 \pm 0.31$ \\
\hline
\hline
\end{tabular}
\end{sidewaystable*}

\end{document}